\documentclass[twocolumn,tighten]{aastex61}
\pdfoutput=1 
\usepackage{amsmath,amstext}
\usepackage{apjfonts} 
\usepackage{afterpage}
\usepackage{xcolor}

\usepackage{graphicx}
\usepackage{epsfig}

\newcommand{\sgra}{Sgr~A$^*$}
\newcommand{\rin}{r_{\rm in}}

\newcommand{\rdif}{r_{\rm dif}}
\newcommand{\rF}{r_{\rm F}}
\newcommand{\rref}{r_{\rm ref}}

\def\gsim{\mathrel{\raise.3ex\hbox{$>$\kern-.75em\lower1ex\hbox{$\sim$}}}}
\def\lsim{\mathrel{\raise.3ex\hbox{$<$\kern-.75em\lower1ex\hbox{$\sim$}}}}

\newcommand{\kzetavM}{k_{\zeta,1}}
\newcommand{\kzetaD}{k_{\zeta,2}}
\newcommand{\kzetaB}{k_{\zeta,3}}

\newcommand{\thetapar}{\theta_{\rm maj}}
\newcommand{\thetaperp}{\theta_{\rm min}}
\newcommand{\thetaparzero}{\theta_{\rm maj,0}}
\newcommand{\thetaperpzero}{\theta_{\rm min,0}}

\setwatermarkfontsize{100px} 

\shorttitle{Scattering model for \sgra}
\shortauthors{Psaltis et al.}

\begin{document}


\title{A Model for Anisotropic Interstellar Scattering and its Application to Sgr~A*}
\shorttitle{Anisotropic Interstellar Scattering}

\author{Dimitrios Psaltis}
\affiliation{Astronomy Department, University of Arizona, 933 N. Cherry Ave, Tucson, AZ 85721}

\affiliation{Radcliffe Institute for Advanced Study, Harvard University, Cambridge, MA 02138, USA}

\affiliation{Black Hole Initiative, Harvard University, 20 Garden St., Cambridge, MA 02138}

\author{Michael D.\ Johnson}
\affiliation{Harvard-Smithsonian Center for Astrophysics, 60 Garden Street, Cambridge, MA 02138, USA}

\affiliation{Black Hole Initiative, Harvard University, 20 Garden St., Cambridge, MA 02138}

\author{Ramesh Narayan}
\affiliation{Harvard-Smithsonian Center for Astrophysics, 60 Garden Street, Cambridge, MA 02138, USA}

\affiliation{Black Hole Initiative, Harvard University, 20 Garden St., Cambridge, MA 02138}

\author{Lia Medeiros}
\affiliation{Astronomy Department, University of Arizona, 933 N. Cherry Ave, Tucson, AZ 85721}
\affiliation{Department of Physics, Broida Hall, University of California Santa Barbara, Santa Barbara, CA 93106}

\author{Lindy Blackburn}
\affiliation{Harvard-Smithsonian Center for Astrophysics, 60 Garden Street, Cambridge, MA 02138, USA}

\author{Geoff C.\ Bower}
\affiliation{Academia Sinica Institute of Astronomy and Astrophysics, 645 N. A{'o}hoku Place, Hilo, HI 96720, USA}


\begin{abstract}
Scattering in the ionized interstellar medium is commonly observed to be anisotropic, with theories of magnetohydrodynamic (MHD) turbulence explaining the anisotropy through a preferred magnetic field direction throughout the scattering regions. In particular, the line of sight to the Galactic Center supermassive black hole, \sgra, exhibits strong and anisotropic scattering, which dominates its observed size at wavelengths of a few millimeters and longer. Therefore, inferences of the intrinsic structure of \sgra\ at these wavelengths are sensitive to the assumed scattering model. In addition, extrapolations of the scattering model from long wavelengths, at which its parameters are usually estimated, to 1.3\,mm, where the Event Horizon Telescope (EHT) seeks to image \sgra\ on Schwarzschild-radius scales, are also sensitive to the assumed scattering model. Past studies of \sgra\ have relied on simple Gaussian models for the scattering kernel that effectively presume an inner scale of turbulence far greater than the diffractive scale; this assumption is likely violated for \sgra\ at 1.3\,mm. We develop a physically motivated model for anisotropic scattering, using a simplified model for MHD turbulence with a finite inner scale and a wandering transverse magnetic field direction. We explore several explicit analytic models for this wandering and derive the expected observational properties --- scatter broadening and refractive scintillation --- for each. For expected values of the inner scale, the scattering kernel for all models is markedly non-Gaussian at 1.3\,mm but is straightforward to calculate and depends only weakly on the assumed model for the wandering of the magnetic field direction. On the other hand, in all models, the refractive substructure depends strongly on the wandering model and may be an important consideration in imaging \sgra\ with the EHT.
\end{abstract}


\section{INTRODUCTION}

The supermassive black hole, \sgra, at the center of the Milky Way lies behind one or more scattering ``screens'' that affect the observational appearance of the black hole at radio wavelengths. Within these scattering regions, small-scale fluctuations in the density of free electrons introduce stochastic phase variations in electromagnetic waves traveling along different paths from the Galactic Center to the Earth. These phase variations lead to a number of astrometric, timing, and spectral effects~\citep[see][for reviews of scattering effects]{Rickett1990, Narayan1992}, the most important of which, in the case of {\sgra}, is the broadening of the source image at millimeter and longer wavelengths.

The properties of the scattering towards \sgra\ have been explored
observationally via measurements of the wavelength dependence of its
scatter-broadened image~\citep[see, e.g.,][]{Lo1993, Lo1998,
  Krichbaum1998,Bower2004, Shen2005,
  Bower2006,Lu2011,Bower2014,Ortiz2016}, via observations of the
images of nearby OH masers and of the free-free emission and
absorption properties of the
screen~\citep{vanLangevelde1992,Frail1994,Lazio1998a,Lazio1998b}, as
well as via measurements of the temporal and angular broadening of the
Galactic Center magnetar~\citep{Spitler2014,Bower2014a}. All these
observations have shown that scattering towards \sgra\ is highly
anisotropic, with a major-to-minor axis ratio for the ensemble average
scattering kernel of $\sim 2$. Recently, \cite{Gwinn2014} discovered
refractive substructure in very long baseline interferometry (VLBI)
observations of \sgra\ at 1.3~cm, demonstrating that, at this
wavelength, scatter-broadening occurs in the "Average" regime (see
Table 1). The location of the scattering screen was originally
inferred based on the scatter broadening of background sources, which
place the screen close to the Galactic Center~\citep{Lazio1998b}. More
recent inferences based on the Galactic Center magnetar place it
closer to Earth, possibly in an intervening spiral
arm~\citep{Bower2014a}. Observations of scattering effects in pulsars
in the direction of the Galactic Center offered hints for a more
complex configuration with multiple screens at various
distances~\citep{Dexter2017}.

Early mm observations~\cite[see, e.g.,][]{Krichbaum1998,
  Doeleman2001}, as well as theoretical expectations of the wavelength
dependence of the image size of the accretion flow around
\sgra~\citep{Ozel2000, Falcke2000}, strongly suggested that, at
mm-wavelengths, the effects of scattering would be minimal.
Indeed, VLBI observations with a small number of baselines have since
demonstrated that the image of \sgra\ at 1.3~mm has a size comparable to
the expected size of the black hole shadow~\citep{Doeleman2008,Fish_2011,Johnson_2015b} 
and that it has asymmetric intrinsic structure~\citep{Fish2016}. At this wavelength,
the inferred source size of ${\sim}40~\mu{\rm as}$ is significantly
larger than the expected scatter broadening (roughly $20\,\mu{\rm as}
\times 10\,\mu{\rm as}$).  Therefore, direct observations of the
shadow that the black hole casts on the surrounding accretion disk
emission~\citep{Bardeen1973,Luminet1979,Falcke2000} should be feasible
and will only be modestly affected by the scattering.

In April 2017, the Event Horizon Telescope (EHT), which is a mm-VLBI array with stations from Hawaii to France and from Arizona to the South Pole, performed the first full-array interferometric observations of \sgra\ with the primary aim of obtaining images with horizon-scale resolution. Mitigating the effects of scattering on these images \citep{Fish2014,Johnson2016}, applying statistical tools to extract the parameters of the black hole shadow~\citep{Psaltis2015} or to compare the observed image to model predictions~\citep[see, e.g.,][]{Broderick2009,Dexter2009, Dexter2010, Moscibrodzka2009, Moscibrodzka2014,Kim2016}, all require a model for the effects of interstellar scattering at EHT wavelengths. The scattering parameters are typically constrained by observations at longer wavelengths~\citep[typically at $\gtrsim 1$~cm; see, e.g.,][]{Bower2006} and a scattering model is used to extrapolate them to EHT wavelengths.

The theory behind the effects of interstellar scattering on the images of astrophysical objects has been developed in a number of studies during the last four decades~\citep[see, e.g.,][]{Cronyn1972, Cohen1974, Lee1975b, Lee1975a, Lee1975, Blandford1985, Goodman1985, Narayan1989a, Goodman1989,Johnson2015,Johnson2016a}. While many of these studies considered isotropic scattering, \cite{Narayan1988} developed a model for anisotropic scattering in planetary atmospheres based on a particular phenomenological model of the anisotropic power spectrum of the underlying density fluctuations; a few other previous studies \citep[e.g.,][]{Goodman1989,Johnson2015} also implemented anisotropy in their discussions. \cite{Chandran2002} studied the effects of interstellar scattering on images using an anisotropic Goldreich-Sridhar power spectrum for the turbulent fluctuations~\citep{Goldreich1995,Goldreich1997} and derived a relation between the degree of anisotropy in the scattering kernel and of the wandering of the projected magnetic field vector along the line-of-sight. 

In fitting data, especially in the case of \sgra, anisotropic
scattering has generally been treated with simple empirical models,
such as an anisotropic Gaussian scattering kernel with a
wavelength-independent orientation and degree of anisotropy (this
model was indeed used in all the observational studies mentioned
earlier). The Gaussian model is adequate so long as the diffractive
scale of scattering is much smaller than the inner scale of turbulence
in the scattering screen. However, in the case of EHT observations, it
is possible that the diffractive scale will become comparable to or
even larger than the turbulence inner scale (see Figure~1 below).  Our aim here is to bridge the gap between the empirical Gaussian kernel approach and more
rigorous theoretical models for scattering.  To this end, we develop a
model of interstellar scattering along the line-of-sight towards
\sgra\ that (i) is anisotropic, (ii) incorporates explicitly the
effects of a finite inner scale, and (iii) has enough degrees of
freedom to model and interpret upcoming EHT data with appropriate
uncertainties in the scattering model incorporated into the error
budget.

\section{The Scattering Screen Towards Sgr A*}

Under the usual assumption of a single, thin scattering screen between
the Earth and \sgra, we can describe fully the statistical properties
of the inhomogeneities that lead to scattering in terms of the phase
structure function $D_\phi(\vec{r})$, where $\vec{r}$ is a transverse
2-vector on the projected scattering screen. The phase structure function
measures the second order correlation between the change in phase $\phi(\vec{r})$
introduced by the screen to the propagating radio waves at two different
locations separated by a transverse distance $\vec{r}$, i.e.,
\begin{equation}
D_\phi(\vec{r})\equiv \langle \left[\phi(\vec{r_0}+\vec{r})-\phi(\vec{r}_0)\right]^2\rangle;.
\end{equation}

This function is related to the phase correlation function $C(\vec{r})$ by
\begin{equation}
D_\phi(\vec{r}) = 2\,[C(0) - C(\vec{r})], \label{Dphir}
\end{equation}
and $C(\vec{r})$ is related to the power spectrum of
fluctuations in the turbulent screen $Q(\vec{q})$ by
\begin{equation}
  C(\vec{r}) = \frac{\lambdabar^2}{4\pi^2}
  \int Q(\vec{q}) \exp(i\vec{q}\cdot\vec{r})
\,d^2q;. \label{Cr}
\end{equation}
Here $\lambda\equiv 2\pi \lambdabar$ is the wavelength of
observations and $\vec{q}$ is a transverse spatial frequency vector.
As written, $D_\phi(\vec{r})$, $C(\vec{r})$ and $Q(\vec{q})$ are all
dimensionless.

We define the diffractive scale $\rdif$ as the transverse length on the scattering screen over which the phase structure function becomes equal to unity, i.e.,
\begin{equation}
D_{\phi}(\rdif)=1\;.
\label{eq:rdif_impl}
\end{equation}
The diffractive scale is related to the angular size of the
scatter-broadened image of a point source $\theta_{\rm scat}$ by 
\begin{equation}
  \rdif\simeq \frac{\sqrt{2\ln 2}}{\pi}\frac{\lambda}{(1+M)\theta_{\rm scat}}\simeq 4.1\times 10^7\left(\frac{\lambda}{\rm cm}\right)^{-1}
  ~{\rm cm}\;.
\label{eq:rdifSGRA}
\end{equation}
The diffractive scale depends on the overall scattering geometry, described using the magnification $M\equiv D/R$, where $D$ is the distance between the screen and the observer, and $R$ is the distance between the screen and the source.
In this work, motivated by observations of \sgra, we take the scattering kernel to be anisotropic. Therefore, $\rdif$ depends on orientation. In the last equality above, we  used typical values of the wavelength-dependent image size of \sgra\ along the major axis of the scattering kernel, as observed at $\sim$cm wavelengths~\citep[see, e.g.,][and references therein]{Bower2006, Psaltis2015}, which lead to the approximation (assuming a Gaussian scattering kernel for simplicity)
\begin{equation}
\theta_{\rm scat}\equiv\theta_{\rm SGRA}\simeq 1.32\left(\frac{\lambda}{{\rm cm}}\right)^2~{\rm mas}.
\label{eq:imageSGRA}
\end{equation}

\begin{figure}[t]
\begin{center}
\includegraphics[width=6cm]{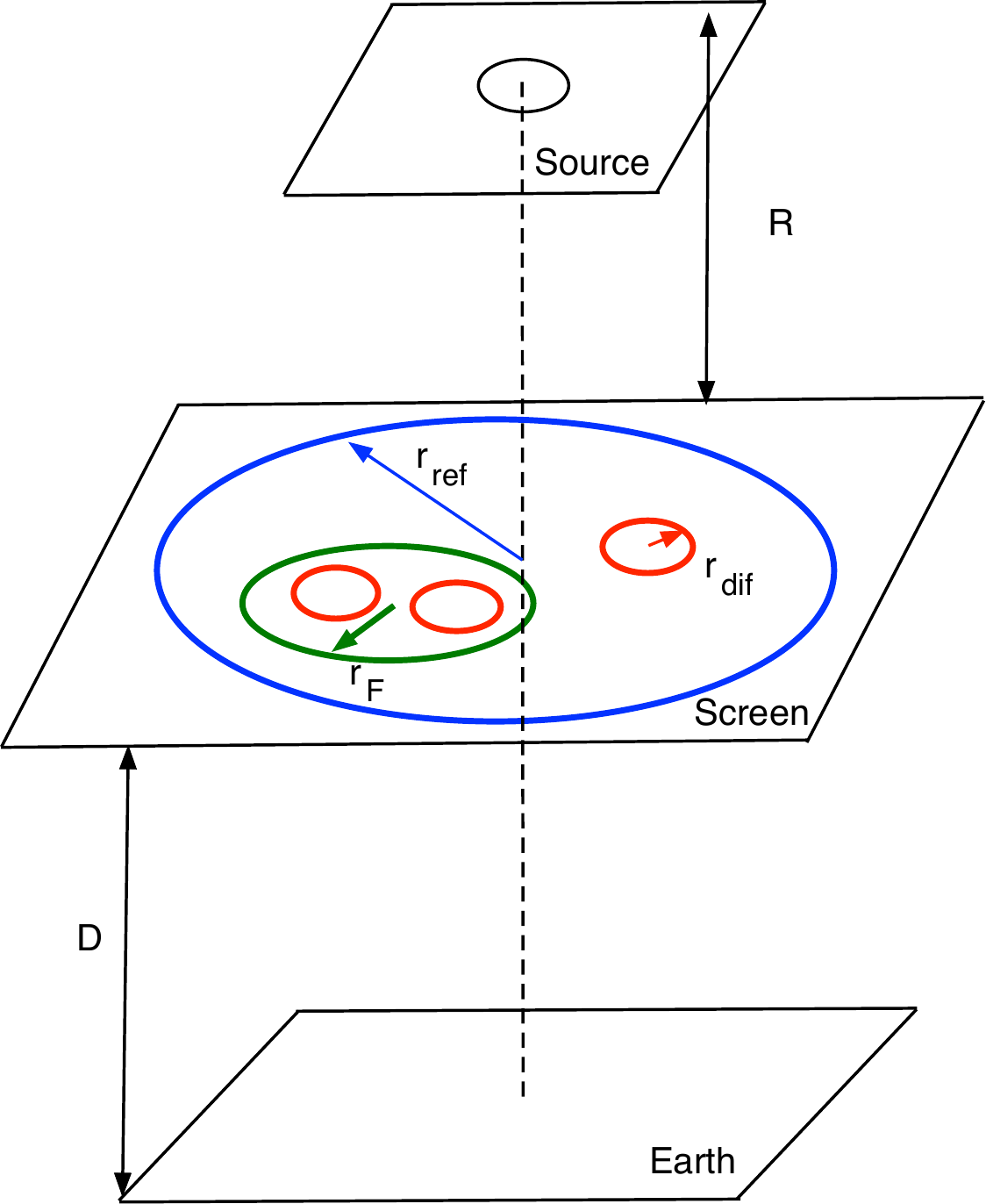}
\caption{A schematic of the diffractive, Fresnel, and refractive scales, as well as a depiction of the definitions of the distances between the Earth, the scattering screen, and the source that we use in the text.
  \label{fig:scales}}
\end{center}
\end{figure}

In addition to the diffractive scale, another scale of importance is the Fresnel scale,
\begin{equation}
\rF\equiv \left(\frac{D R}{D+R}\lambdabar\right)^{1/2}=\left(\frac{D\lambdabar}{1+M}\right)^{1/2}\;,
\end{equation}
where $\lambdabar=\lambda/2\pi$ is the wavelength of observations.

In the strong scattering regime, defined as the regime where $\rdif\ll \rF$, a third important scale is the refractive scale,
\begin{equation}
\rref\equiv\frac{\rF^2}{\rdif}\;,
\end{equation}
which corresponds to the projected size of the image of a point source on the scattering screen, i.e.,
\begin{equation}
\rref\simeq \frac{\theta_{\rm scat} D}{2 \sqrt{2 \ln 2}}\simeq 2.3\times 10^{13}
\left(\frac{\lambda}{{\rm cm}}\right)^2\left(\frac{D}{2.7~{\rm kpc}}\right)~{\rm cm}\;,
\label{eq:rrefSGRA}
\end{equation}
where we have used the distance to the screen as inferred from observations of the Galactic Center magnetar~\citep{Bower2014a}. All three length scales, $\rdif$, $\rF$, $\rref$, depend on wavelength. 

\begin{figure}[t]
\begin{center}
\includegraphics[height=7cm,width=8cm]{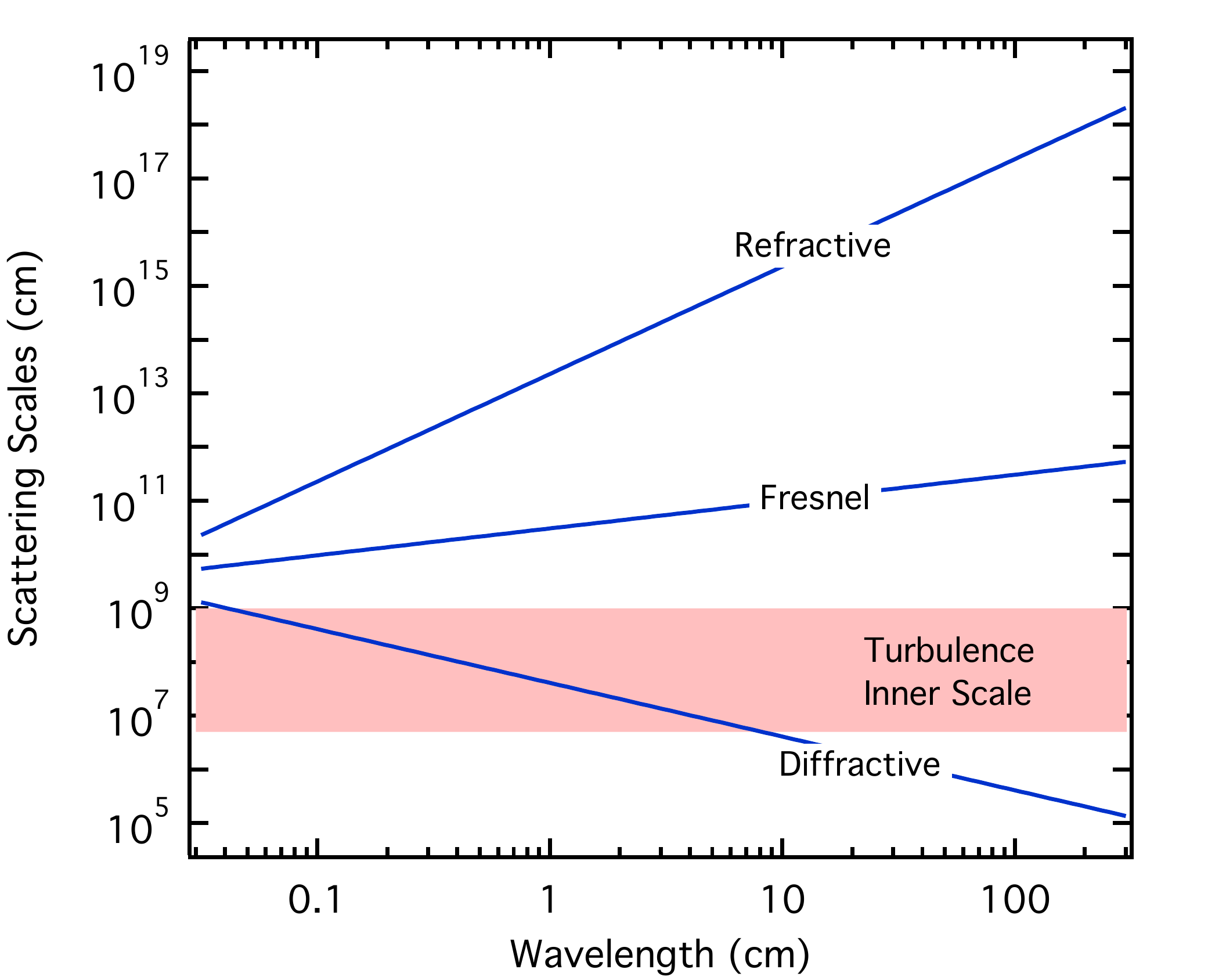}
\caption{The diffractive, Fresnel, and refractive scales as a function
  of wavelength for the typical parameters of the screen towards
  \sgra\ and for the magnification set to zero. The horizontal shaded
  band shows the range of inner scale of turbulence, as inferred for
  the interstellar medium in general and for the screen towards
  \sgra\ in particular. At the $\simeq$mm wavelengths of the EHT, it
  is likely that the diffractive scale is comparable to or larger than
  the inner scale of turbulence.
  \label{fig:scater_scales}}
\end{center}
\end{figure}

Figures~\ref{fig:scales} and \ref{fig:scater_scales} show a schematic diagram and the sizes of the three characteristic scales for the scattering screen towards \sgra\ as a function of wavelength, respectively. The second figure also shows estimates of the inner scale of turbulence in the scattering screen towards \sgra, which range from $\simeq 50-10^4$~km \citep[see, e.g.,][]{Spangler1990,Rickett_2009,Gwinn2014}. At the $\sim 3-20$~cm wavelengths where measurements of the properties of the scattering screen towards \sgra\ are usually carried out, the inner scale of turbulence is larger than the diffractive scale, consistent with the measurement of a quadratic dependence of the scatter-broadened image size on wavelength~\citep[see, e.g.,][]{Bower2004, Bower2006}. However, it is likely that the diffractive scale becomes larger than the inner scale of turbulence at the $\simeq$mm wavelengths of the EHT. As a result, extrapolating the properties of the scattering screen from a few cm to $\simeq$mm wavelengths requires taking explicitly into account the finite size of the inner scale of turbulence in the screen.

\begin{deluxetable*}{lccc}
\tablewidth{0pt}
\tablecolumns{4}
\tablecaption{Regimes of interferometric imaging of scatter broadening~\citep[after][]{Narayan1989a, Goodman1989}.\label{table:regimes}}
\tablehead{& \colhead{Ensemble Average} & \colhead{Average}
& \colhead{Snapshot}}
\startdata
Integration Time & $t_{\rm int} \gg t_{\rm ref}$ & 
$t_{\rm ref}\gg t_{\rm int} \gg t_{\rm dif}$ & $t_{\rm dif}\gg t_{\rm int}$\\
Bandwidth & $\frac{\Delta\nu}{\nu}\gg 1$ & $1\gg\frac{\Delta\nu}{\nu}\gg 
\frac{\rdif^2}{\rref^2}$ & $\frac{\rdif^2}{\rref^2}\gg\frac{\Delta\nu}{\nu}$\\
Source Size & \nodata & $\theta_{\rm src}\lsim \frac{\lambda}{\rdif}\simeq
\theta_{\rm scat}$ & $\theta_{\rm src}\lsim \frac{\rdif}{D}\simeq
\left(\frac{\rdif}{\rref}\right)\theta_{\rm scat}$ \\
\enddata
\end{deluxetable*}

The various characteristic length scales allow us to define two corresponding timescales: the diffractive timescale
\begin{equation}
t_{\rm dif}\equiv \frac{r_{\rm dif}}{u}
\simeq 8.2\left(\frac{\lambda}{\rm cm}\right)^{-1}
\left(\frac{u}{50~\mbox{km~s}^{-1}}\right)^{-1}~{\rm s}\;,
\label{eq:tdif}
\end{equation}
and the refractive timescale
\begin{equation}
t_{\rm ref}\equiv \frac{r_{\rm ref}}{u}\simeq
0.15
\left(\frac{\lambda}{{\rm cm}}\right)^2\left(\frac{D}{2.7~{\rm kpc}}\right)
\left(\frac{u}{50~\mbox{km~s}^{-1}}\right)^{-1}~{\rm yr}\;,
\label{eq:tref}
\end{equation}
where $u$ is the transverse relative velocity between the source, the screen, and the observer. \cite{Narayan1989a} and \cite{Goodman1989} identified three regimes of scattering effects on interferometric imaging: the snapshot regime, the average regime, and the ensemble average regime (see above references for the exact definitions). The regime applicable in each case depends on the integration time $t_{\rm int}$ and the bandwidth $\delta \nu/\nu$ of the instrument used for the measurement of the interferometric visibilities, as well as on the source size (before scattering) $\theta_{\rm src}$. Table~\ref{table:regimes} summarizes the conditions for the three regimes~\citep[based on the work of][]{Goodman1989} under the assumption that the characteristic baseline length at a given wavelength is comparable to $\rdif$. For all wavelengths and instruments of interest, interferometric imaging of \sgra\ takes place in the average regime~\citep[see also][]{Gwinn2014, Johnson2015, Johnson2016a, Johnson2016}. 

\section{Modeling the source of anisotropy}

Interstellar scattering is caused by electron density fluctuations
along the line-of-sight, which are believed to be the result of
turbulence in the magnetized interstellar medium. Models of MHD
turbulence \citep{Goldreich1995} indicate that turbulent eddies are
highly elongated along the local direction of the magnetic field. This
anisotropy in density fluctuations should induce anisotropic
scattering. As discussed in the introduction, high resolution
observations of \sgra\ at cm wavelengths do reveal anisotropically
scatter-broadened images, with major and minor axes in a two-to-one
ratio. The scattering angles along the two axes scale as the square of
the observing wavelength~\citep[see, e.g.,][]{Bower2004, Bower2006}; a
priori, this seems inconsistent with a turbulent spectrum of density
fluctuations, which would predict a scaling proportional to
$\lambda^{11/5}$ for a Kolmogorov-like turbulence spectrum
\citep[e.g.,][]{Goodman1985}. However, a quadratic wavelength
dependence would be consistent with expectations, if the inner scale
of the turbulence is larger than the diffractive scale, as is expected to be 
the case at cm and longer wavelengths
(Fig.~\ref{fig:scater_scales}). However, we expect $\rdif$ to be
comparable to the inner scale at the shorter (mm) wavelengths of the
EHT, and hence there is a need to go beyond the usual $\lambda^2$
models of scattering.

Motivated by the above, we include the following features in our model of the phase structure function $D_\phi(\vec{r})$ and its corresponding power spectrum $Q(\vec{q})$:

\noindent {\em (i)\/} The power spectrum has a power-law form $\propto q^{-(\alpha+2)}$, where the index $\alpha$ is a free parameter but is expected to be
close to the Kolmogorov value $\alpha = 5/3$.

\noindent {\em (ii)\/} The power spectrum has an inner scale, which we will call $\rin$.

\noindent {\em (iii)\/} The scattering is anisotropic, reflecting the presence of a preferred
direction of the projected magnetic field.

\noindent {\em (iv)\/} The strength of the scattering and the degree and orientation of its anisotropy vary with depth in the scattering screen. To accommodate this, we decompose the screen into a superposition of independent layers along the line-of-sight (though still within a single thin screen approximation).

\noindent {\em (v)\/} Within any given layer in the screen, we assume
infinite anisotropy of the relevant turbulent eddies. According to the
\citet{Goldreich1995} theory of MHD turbulence, anisotropic eddies on
a given scale $r$ have an axis ratio $\sim(r/L_{\rm out})^{1/3}$,
where $L_{\rm out}$ is the outer scale of the turbulence. \citet[][see
  also \citealt{Armstrong1981}]{Armstrong1995} estimated that $L_{\rm
  out}$ is definitely greater than $10^{15}$\,cm, and likely $>
10^{20}$\,cm, whereas the largest scale $r$ of interest for
observations of \sgra\ is $\sim10^{9}$\,cm. Thus, eddies on the scales
of interest have axis ratios of at least 100:1, which we may safely
model as infinitely anisotropic.

Making use of the above assumptions, we write the differential
contribution to the power spectrum of fluctuations from a layer at
depth $z$ along the line-of-sight as
\begin{equation}
\frac{dQ(\vec{q})}{dz} = Q_z\, (q\rin)^{-(\alpha+2)} \exp(-q^2\rin^2)\,
\delta(\phi_q-\phi_z)\;. \label{dQdz}
\end{equation}
Here, the (two-dimensional) transverse wave-vector $\vec{q}$ has been written in polar form, with $q$
representing the magnitude of $\vec{q}$ and $\phi_q$ its orientation, $Q_z$
measures the strength of the scattering per unit distance along the
line-of-sight, and $\phi_z$ indicates the direction along which the
anisotropic phase fluctuations are oriented. Given the nature of MHD
turbulence, $\phi_z$ is oriented orthogonal to the projected
direction of the magnetic field in the layer under consideration. 

We assume that all layers in the scattering screen can be described by
the same spectral index $\alpha$ and inner scale $\rin$, because all
infinitesimal layers are different realizations of the same underlying
turbulence. The only two properties that we allow to vary between
different layers are the amplitude of fluctuations $Q_z$ and the
orientation of the anisotropy $\phi_z$. The delta-function in
equation~(\ref{dQdz}) indicates that we take the anisotropy within
each layer to be infinite.

Substituting equation~(\ref{dQdz}) into equations~(\ref{Dphir}) and (\ref{Cr}), and writing
$\vec{r}$ in polar coordinates $(r,\phi)$, we obtain
\begin{eqnarray}
\frac{dD_\phi(\vec{r})}{dz} &=& \frac{\lambdabar^2 Q_z}{2\pi^2} \int
(q\rin)^{-(\alpha+2)}
\exp(-q^2\rin^2)\,\left[1-\cos(\vec{q}\cdot\vec{r})\right]\,\nonumber\\
&&\qquad\delta(\phi_q-\phi_z)\,d^2q \nonumber \\ 
&=& \frac{\lambdabar^2 Q_z}{2 \pi^2\rin^2} \int_0^\infty
q^{\prime-(\alpha+1)}\exp(-q^{\prime 2})\nonumber\\
&&\qquad\left[1-\cos\left(q^\prime\frac{r}{\rin}\cos(\phi-\phi_z)\right)\right]\,dq^\prime
\nonumber\\
& =& \frac{4 {\cal C}}{\alpha}
\left[M\left(-\frac{\alpha}{2},\,
  \frac{1}{2},\,
  -\frac{r^2}{4\rin^2}\cos^2(\phi-\phi_z)\right)-1\right],
\label{eq:dDdz}
\end{eqnarray}
where $q^\prime\equiv q \rin$, $M$ is the Kummer (or confluent hypergeometric) function and we have defined the coefficient $\cal{C}$ to be
\begin{equation}
{\cal C} \equiv \frac{\lambdabar^2 Q_z
  \Gamma\left(1-\frac{\alpha}{2}\right)} {8 \pi^2 \rin^2}\;.
  \label{eq:calC}
\end{equation}

In order to calculate the phase structure function for the scattering
screen, we need to integrate equation~(\ref{eq:dDdz}) over the
line-of-sight depth $z$ of the screen, i.e.,
\begin{equation}
  D_\phi(\vec{r})=\int_0^{z_0} \frac{dD_\phi(\vec{r})}{dz} dz \;.
\end{equation}
In writing this integral, we assume that both the strength of
perturbations per unit distance, $Q_z$, and the orientation of
anisotropy $\phi_z$ vary with the line-of-sight coordinate $z$.  In
principle, as the magnetic field orientation wanders along the
line-of-sight, those two functions may be non-monotonic.  However,
we can rewrite the integral in a more tractable form by introducing an
appropriately averaged amplitude $\bar{Q}_z$ in the definition~(\ref{eq:calC}) of ${\cal C}$ 
and a normalized probability density $P(\phi_z)$ that measures the relative frequency
of layers in the scattering screen that are anisotropic along a
particular orientation $\phi_z$. With these definitions, we can write
the phase structure function as
\begin{equation}
  D_\phi(\vec{r})=\int_0^{2\pi} \frac{dD_\phi(\vec{r})}{dz}
  P(\phi_z) d\phi_z \;.
  \label{eq:Dphi_int}
\end{equation}
Equivalently, we can write the integrated power spectrum of perturbations
along the line-of-sight as
\begin{equation}
  Q(\vec{q})=\int_0^{2\pi} \frac{dQ(\vec{q})}{dz}
  P(\phi_z) d\phi_z \;.
  \label{eq:Q_int}
\end{equation}

The integral in (\ref{eq:Dphi_int}) is difficult to evaluate
analytically even for simple forms of the probability distribution
function. Analytical results are possible in certain asymptotic
limits, which we now discuss. Calculating these limits allows us, in
the following section, to write the anisotropic scattering model in
terms of quantities that are useful in fitting observational data.

\subsection{The quadratic limit $r\ll \rin$}

We first consider the limit of small transverse distances, $r \ll
 \rin$, and use the series expansion
\begin{equation}
M(a,b,z) = 1 +\frac{a}{b}\,z + \frac{a(a+1)}{b(b+1)}\,\frac{z^2}{2!}\,
+\, ...\,,
\end{equation}
to obtain
\begin{equation}
  D_{\phi}(\vec{r})\simeq {\cal C} \left[
  \int_{0}^{2\pi}\cos^2(\phi-\phi_z)
  P(\phi_z-\phi_0)d\phi_z\right]
  \frac{r^2}{\rin^2}\;. \label{eq:smallr}
\end{equation}
In this last expression, we wrote the probability distribution function
in a general form that encaptulates the fact that we will choose below
models that are peaked and symmetric around a prefered orientation at
an angle $\phi_0$, i.e.,
\begin{equation}
  P(\phi_z-\phi_0)=P(\phi_0-\phi_z)\;.
\end{equation}
They will also be point symmetric around the origin, i.e.,
\begin{equation}
  P(\phi_z-\phi_0)=P(\phi_z-\phi_0+\pi)
\end{equation}
in order for the phase structure function (being the Fourier transform
of the power spectrum) to be real.

We can use these assumed properties of the probability distribution function
to write equation~(\ref{eq:smallr}) (after making the change of variables
$\phi_z=\phi_0+\phi^\prime$) as
\begin{eqnarray}
  D_{\phi}(\vec{r})&=& {\cal C} \left[
  \int_{0}^{2\pi}\frac{1+\cos(2\phi-2\phi_z)}{2}
  P(\phi_z-\phi_0)d\phi_z\right]
  \frac{r^2}{\rin^2}\nonumber\\
  &=& \frac{{\cal C}}{2}\left\{
  1+\cos(2\phi-2\phi_0)\int_0^{2\pi}\cos(2\phi^\prime)P(\phi^\prime)d\phi^\prime
  \right.\nonumber\\
&&\quad \left.
  +\sin(2\phi-2\phi_0)\int_0^{2\pi}\sin(2\phi^\prime)P(\phi^\prime)d\phi^\prime\right\}
  \frac{r^2}{\rin^2}\nonumber\\
    &=& \frac{{\cal C}}{2}\left\{
  1+\zeta_0\cos[2(\phi-\phi_0)]  \right\}  \frac{r^2}{\rin^2}
  \label{eq:Dsmallr}
\end{eqnarray}
where
\begin{equation}
  \zeta_0\equiv\int_0^{2\pi}\cos(2\phi_z)P(\phi_z)d\phi_z\;.
  \label{eq:zeta_general}
\end{equation}
As expected, the structure function in this limit increases
quadratically with $r$ and has the characteristic angular dependence
given in equation~(\ref{eq:Dsmallr}). In this limit, the form of the
phase structure function does not depend on the particular model for the
wandering of the magnetic field. The latter only determines the degree of
anisotropy via equation~(\ref{eq:zeta_general}).

For simplicity in our notation, we will denote the phase structure
function along the orientation of anisotropy $\phi_0$ (i.e., the major
axis) by
\begin{equation}
  D_{\rm maj}(r)\equiv D_\phi(r,\phi=\phi_0)
  = \frac{{\cal C}(1+\zeta_0)}{2} \frac{r^2}{\rin^2}\;,
  \label{eq:Dmaj_smallr}
\end{equation}
and perpendicular to this orientation by
\begin{equation}
  D_{\rm min}(r)\equiv D_\phi(r,\phi=\phi_0+\pi/2)
  = \frac{{\cal C}(1-\zeta_0)}{2} \frac{r^2}{\rin^2}\;.
  \label{eq:Dmin_smallr}
\end{equation}
Clearly, in this limit
\begin{equation}
  D_{\rm maj}(r)+D_{\rm min}(r)={\cal C}\frac{r^2}{\rin^2}\;.
\end{equation}

\subsection{The limit $r\gg \rin$}

In the opposite limit of large transverse distances, $r \gg \rin$, 
we use the corresponding asymptotic expansion
\begin{equation} M(a,b,z) \approx
  \frac{\Gamma(b)\,\exp(z)\,z^{(a-b)}}{\Gamma(a)}\, +\,
  \frac{\Gamma(b)\,(-z)^{-a}}{\Gamma(b-a)}.
\end{equation}
The second term dominates, so we find
\begin{eqnarray}
D_\phi(\vec{r}) &\approx& \frac{2^{2-\alpha}\sqrt{\pi}{\cal C}}
{\alpha\Gamma[(1+\alpha)/2]}\nonumber\\
&&\quad
 \left[\int_0^{2\pi}\left\vert\cos(\phi-\phi_z)\right\vert^\alpha
  P(\phi_z-\phi_0)d\phi_z\right]
     \left(\frac{r}{\rin}\right)^\alpha\;.
\label{eq:larger}
\end{eqnarray}
In this limit, the structure function increases as $r^\alpha$ rather
than as $r^2$.

The functional form of the above integral depends sensitively on the
functional form of the model for the wandering of the magnetic field.
However, for reasons that will become apparent below, we will write
the phase structure function along the orientation of anisotropy
$\phi_0$ as
\begin{eqnarray}
    D_{\rm maj}(r)&=&D_\phi(r,\phi=\phi_0) =
    \frac{2^{2-\alpha}\sqrt{\pi}{\cal C}}
{\alpha\Gamma[(1+\alpha)/2]}\nonumber\\
&&
 \left[\int_0^{2\pi}\left\vert\cos(\phi_0-\phi_z)\right\vert^\alpha
  P(\phi_z-\phi_0)d\phi_z\right]
 \left(\frac{r}{\rin}\right)^\alpha\nonumber\\
 &=&
 \frac{{\cal C}(1+\zeta_0)}{2}{\cal B}_{\rm maj}
 \left(\frac{r}{\rin}\right)^\alpha\;,
  \label{eq:Dmaj_larger}
\end{eqnarray}
where $\zeta_0$ is given by equation~(\ref{eq:zeta_general}) and
\begin{eqnarray}
  {\cal B}_{\rm maj}&\equiv&
      \frac{2^{3-\alpha}\sqrt{\pi}}
{\alpha\Gamma[(1+\alpha)/2](1+\zeta_0)}\nonumber\\
&&\qquad
 \left[\int_0^{2\pi}\left\vert\cos(\phi_0-\phi_z)\right\vert^\alpha
  P(\phi_z-\phi_0)d\phi_z\right]\;.
\label{eq:B_maj}
\end{eqnarray}
Similarly, we write the phase structure function perpendicular to
the orientation of anisotropy as
\begin{eqnarray}
    D_{\rm min}(r)&=&D_\phi(r,\phi=\phi_0+\pi/2) =
    \frac{2^{2-\alpha}\sqrt{\pi}{\cal C}}
{\alpha\Gamma[(1+\alpha)/2]}\nonumber\\
&&
 \left[\int_0^{2\pi}\left\vert\sin(\phi_0-\phi_z)\right\vert^\alpha
  P(\phi_z-\phi_0)d\phi_z\right]
 \left(\frac{r}{\rin}\right)^\alpha\nonumber\\
 &=&
 \frac{{\cal C}(1-\zeta_0)}{2}{\cal B}_{\rm min}
 \left(\frac{r}{\rin}\right)^\alpha\;,
  \label{eq:Dmin_larger}
\end{eqnarray}
where
\begin{eqnarray}
  {\cal B}_{\rm min}&\equiv&
      \frac{2^{3-\alpha}\sqrt{\pi}}
{\alpha\Gamma[(1+\alpha)/2](1-\zeta_0)}\nonumber\\
&&\qquad
 \left[\int_0^{2\pi}\left\vert\sin(\phi_0-\phi_z)\right\vert^\alpha
  P(\phi_z-\phi_0)d\phi_z\right]\;.
\label{eq:B_min}
\end{eqnarray}

\subsection{The General Case}

In order to facilitate efficient and accurate calculations of the
phase structure function in the general case, we devised an
approximate analytic expression that bridges the two limiting cases
discussed above. We start by writing an approximate expression for the
phase structure function along the major axis of anisotropy as
\begin{eqnarray}
  D_{\rm maj}(r)&=&\frac{{\cal C}(1+\zeta_0)}{2}B_{\rm maj}
  \left(\frac{2}{\alpha B_{\rm maj}}\right)^{-\alpha/(2-\alpha)}
  \nonumber\\
  &&
  \left\{\left[1+\left(\frac{2}{\alpha B_{\rm maj}}\right)^{2/(2-\alpha)}
    \left(\frac{r}{\rin}\right)^2\right]^{\alpha/2}-1\right\}
\end{eqnarray}
and perpendicular to it as
\begin{eqnarray}
  D_{\rm min}(r)&=&\frac{{\cal C}(1-\zeta_0)}{2}B_{\rm min}
  \left(\frac{2}{\alpha B_{\rm min}}\right)^{-\alpha/(2-\alpha)}
  \nonumber\\
  &&
  \left\{\left[1+\left(\frac{2}{\alpha B_{\rm min}}\right)^{2/(2-\alpha)}
    \left(\frac{r}{\rin}\right)^2\right]^{\alpha/2}-1\right\}
\end{eqnarray}

Using these two approximate expressions, we write the general, approximate
form for the phase structure function as
\begin{eqnarray}
  D_\phi(r,\phi)&=&\left[\frac{D_{\rm maj}(r)+D_{\rm min}(r)}{2}\right]
  \nonumber\\
  &&\qquad
  +\left[\frac{D_{\rm maj}(r)-D_{\rm min}(r)}{2}\right]\cos[2(\phi-\phi_0)]\;.
  \label{eq:D_general}
\end{eqnarray}
Numerical tests show that this fitting function introduces an error no
larger than 2\% for any value of the ratio $r/\rin$, for values of the
power-law index $\alpha$ lying in the physicaly interesting range
between $\alpha=5/3$ (Kolmogorov) and $\alpha=2$, and for the various
models of the wandering of the magnetic field that we introduce below.

\section{Probabilistic Models for the Wandering of the Direction of the Magnetic Field}

In this section, we describe three different analytic, probabilistic
models for the wandering of the direction of the magnetic field along
the line of sight in the scattering screen.

\subsection{von Mises Model}

For most of the calculations below, we use the amphidirectional von
Mises distribution~\citep{Mardia1999} centered at orientation $\phi_0$
and with concentration parameter $\kzetavM$,
\begin{eqnarray}
  P_{\rm vM}(\phi_z,\phi_0) &=& \frac{1}{4\pi I_0(\kzetavM)}
  \left[e^{\kzetavM\cos(\phi_z-\phi_0)}
    +e^{\kzetavM\cos(\phi_z-\phi_0+\pi)}\right] \nonumber \\
    &=& \frac{1}{2\pi I_0(\kzetavM)}\ \cosh\,[\kzetavM \cos(\phi_z-\phi_0)]\;,
    \label{eq:vonMis}
\end{eqnarray}
where $I_0(k_{\zeta,1})$ is the modified Bessel function of the first kind
and order zero and the denominator is needed to normalize the
probability distribution.  The von Mises function is a generalization
of the Gaussian distribution for circular quantities such as angles,
with the concentration parameter $\kzetavM$ providing a measure of the
"peakiness" of the distribution. Note that the power
spectrum~(\ref{eq:Q_int}) and the probability distribution
(\ref{eq:vonMis}) need to be point-symmetric around the origin, in
order to describe a real distribution of density
inhomogeneities. Figure~{\ref{fig:vonMis} shows the amphidirectional
  von Mises distribution for different values of the concentration
  parameter. 
  
  For this model, equation~(\ref{eq:zeta_general}) gives
\begin{equation}
  \zeta_0=\frac{I_2(k_{\zeta,1})}{I_0(k_{\zeta,1})}\;.
\end{equation}
As we discuss below, for \sgra, $\zeta_0\simeq 3/5$ and plugging this value into the above equation gives $k_{\zeta,1}\simeq 4.38$. Similarly, equation~(\ref{eq:B_maj}) gives
\begin{equation}
  {\cal B}_{\rm maj}=\frac{2^{3-\alpha}\sqrt{\pi}}
  {\alpha[I_0(\kzetavM)+I_2(\kzetavM)]}
  \;_1\tilde{F}_2\left(\frac{1+\alpha}{2};\frac{1}{2},1+\frac{\alpha}{2};\frac{\kzetavM^2}{4}\right)\;.
\end{equation}
where $\;_1\tilde{F}_2$ is the regularized hypergeometric function, and equation~(\ref{eq:B_min}) gives
\begin{equation}
  {\cal B}_{\rm min}=\frac{2^{2-\alpha} k_{\zeta,1}}
  {\alpha I_1(k_{\zeta,1})}
   \;_0\tilde{F}_1\left(1+\frac{\alpha}{2},\frac{\kzetavM^2}{4}\right)\;.
\end{equation}
Both coefficients ${\cal B}_{\rm maj}$ and ${\cal B}_{\rm min}$ are equal to unity at $\alpha=2$ and, for small values of $\kzetavM$, they scale approximately as $\alpha^{-2}$ down to a value comparable to 2 at $\alpha=1$. For a Kolmogorov spectrum (i.e., $\alpha=5/3$) and for the value of the concentration parameter $\kzetavM$ that we inferred above for \sgra, we find ${\cal B}_{\rm maj}\simeq 1.54$ and ${\cal B}_{\rm min}\simeq 1.79$.

\begin{figure}[t]
\begin{center}
\includegraphics[height=7cm,width=8cm]{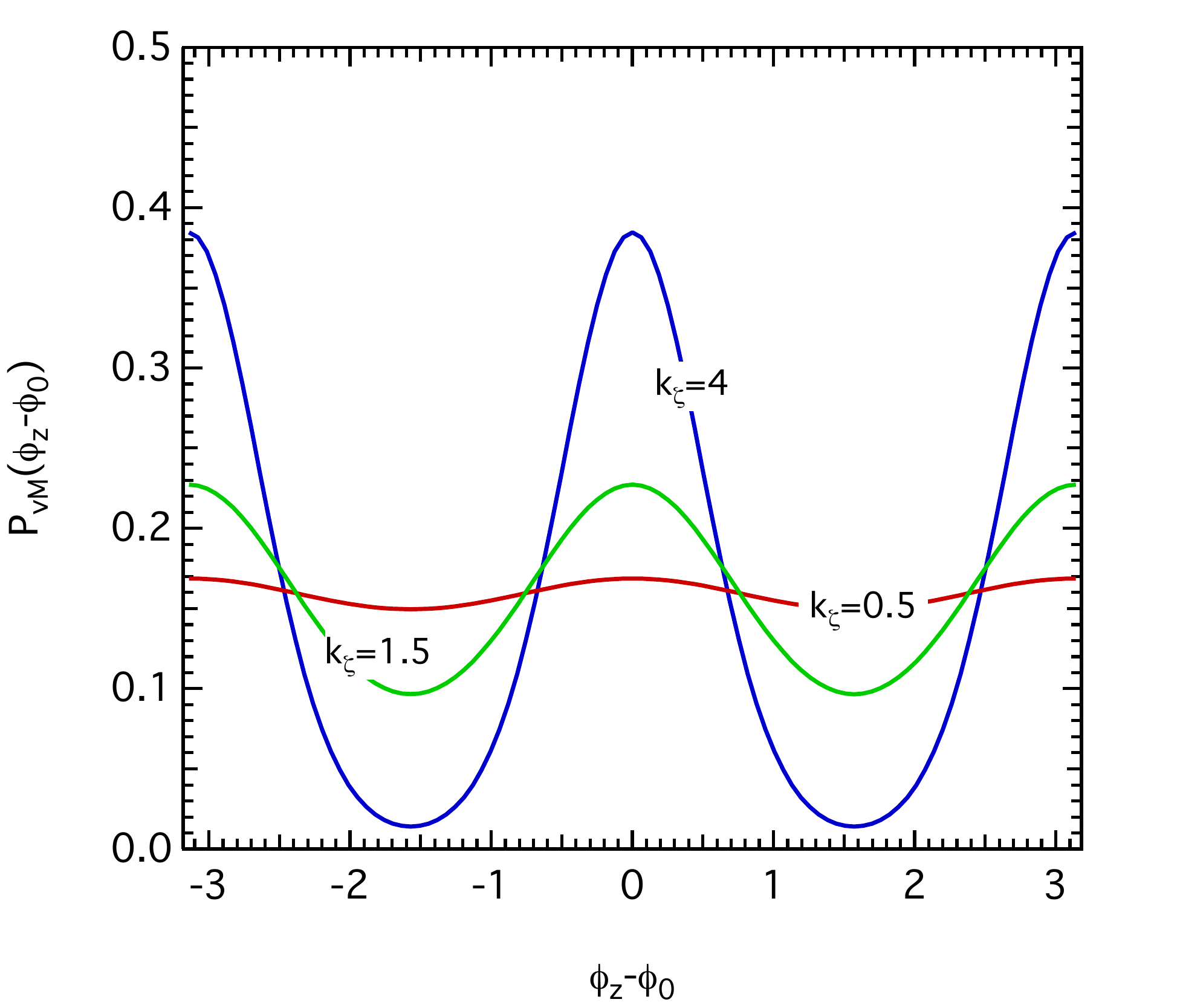}
\caption{The amphidirectional von Mises distribution for different
  values of the concentration parameter $k_{\zeta,1}$. The 
  asymmetries of the resulting scattering kernels that correspond to $k_\zeta=0.5$,
  $1.5$, and $4$ are ${\cal A}=1.03$, 1.23, and 1.91, respectively.
  \label{fig:vonMis}}
\end{center}
\end{figure}

\subsection{Dipole model}

As a second model, we use the anisotropic scattering model introduced
in~\cite{Johnson2016a}, which we refer to hereafter as the ``dipole''
model,
\begin{equation}
  P_2(\phi_z-\phi_0; k_{\zeta,2})=\frac{\left[1+k_{\rm \zeta,2}\sin^2(\phi_z-\phi_0)\right]^{-1-\alpha/2}}{2\pi \;_2F_1\left(1/2,1+\alpha/2,1,-k_{\rm \zeta,2}\right)}\;.
    \label{eq:model2}
\end{equation}
This expression arises from equation~(6) of \cite{Johnson2016a} that
is written in Cartesian coordinates, after making the identification
\begin{equation}
  k_{\rm \zeta,2}\equiv\frac{r_{0y}^2}{r_{0x}^2}-1
\end{equation}
and integrating the resulting expression to obtain the normalization
constant that appears in the denominator of
equation~(\ref{eq:model2}). 

For this model, equation~(\ref{eq:zeta_general}) gives
\begin{equation}
    \zeta_0= \frac{{\cal A}^2-1}{{\cal A}^2+1}\;,
\end{equation}
where the quantity
\begin{eqnarray}
 {\cal A}^2&=& 
\frac{
  \,_2F_1\left(\frac{\alpha+2}{2},\frac{1}{2},2,-\kzetaD\right)
}{
  \,_2F_1\left(\frac{\alpha+2}{2},\frac{3}{2},2,-\kzetaD\right)\;
  }
\end{eqnarray}
is related to the asymmetry of the scattering kernel (see \S5.1
below).  For \sgra, $\zeta_0\simeq 3/5$ and, therefore, $\kzetaD\simeq
3.52$.  Furthermore, equation~(\ref{eq:B_maj}) gives
\begin{eqnarray}
  {\cal B}_{\rm maj}&=&\frac{2^{4-\alpha}}
  {\alpha^2(1+\zeta_0)\Gamma(\alpha/2)\sqrt{1+\kzetaD}}\nonumber\\
  &&\qquad\qquad
    \times\,_2{F}_1\left(\frac{1}{2},\frac{2+\alpha}{2},1,-k_{\zeta,2}\right)^{-1}\;,
\end{eqnarray}
and equation~(\ref{eq:B_min}) gives
\begin{eqnarray}
  {\cal B}_{\rm min}&=&\frac{2^{4-\alpha}}
  {\alpha^2(1-\zeta_0)\Gamma(\alpha/2)}\nonumber\\
  &&\qquad\qquad
    \times\,_2{F}_1\left(\frac{1}{2},-\frac{\alpha}{2},1,-k_{\zeta,2}\right)^{-1}\;,
\end{eqnarray}
For a Kolmogorov spectrum (i.e., $\alpha=5/3$) and for the value of the concentration parameter $\kzetaD$ that we inferred above for \sgra, we find ${\cal B}_{\rm maj}\simeq 1.54$ and ${\cal B}_{\rm min}\simeq 1.75$.
  
\subsection{Periodic Boxcar Model}

Finally, we also consider a periodic boxcar model of the form
\begin{eqnarray}
&&  P_3(\phi_z-\phi_0; k_{\zeta,3})=\frac{1+k_{\zeta,3}}{2\pi}
  \nonumber\\
&&  \times
  \left\{
  \begin{array}{ll}
  1, & {\rm if}~(1+k_{\zeta,3})\vert\phi_z-\phi_0+n\pi\vert\le \pi/2\;;\quad n=...,-1,0,1,...\\
  0, & {\rm otherwise}\\
  \end{array}
  \right.   \nonumber\\
  &&\label{eq:model3}
\end{eqnarray}
with the concentration parameter $k_{\zeta,3}$ defined in a way
analogous to the earlier two models, i.e., $k_{\zeta,3}=0$ corresponds
to an isotropic angular model and the degree of anisotropy increases
with $k_{\zeta,3}$.

For this model, equation~(\ref{eq:zeta_general}) gives
\begin{equation}
  \zeta_0=\frac{1+k_{\zeta,3}}{\pi}
  \sin\left(\frac{\pi k_{\zeta,3}}{1+k_{\zeta,3}}\right)
\end{equation}
For \sgra, $\zeta_0=3/5$ and, therefore, $k_{\zeta,3}\simeq 0.89$.
Equations~(\ref{eq:B_maj}) and (\ref{eq:B_min}) can be also be
integrated analytically in terms of hypergeometric functions. However,
the expressions are long and unwieldy. As a result, these expressions
are best evaluated numerically. For a Kolmogorov spectrum (i.e., $\alpha=5/3$) and for the value of the concentration parameter $\kzetaB$ that we inferred above for \sgra, we find ${\cal B}_{\rm maj}\simeq 1.55$ and ${\cal B}_{\rm min}\simeq 1.85$.

Figure~\ref{fig:three_models} compares the detailed functional form of
the three different angular models for values of the concentration
parameters that give rise to a two-to-one anisotropy in the scattering
kernel at very long wavelengths (see below).

\begin{figure}[t]
\begin{center}
\includegraphics[height=7cm,width=8cm]{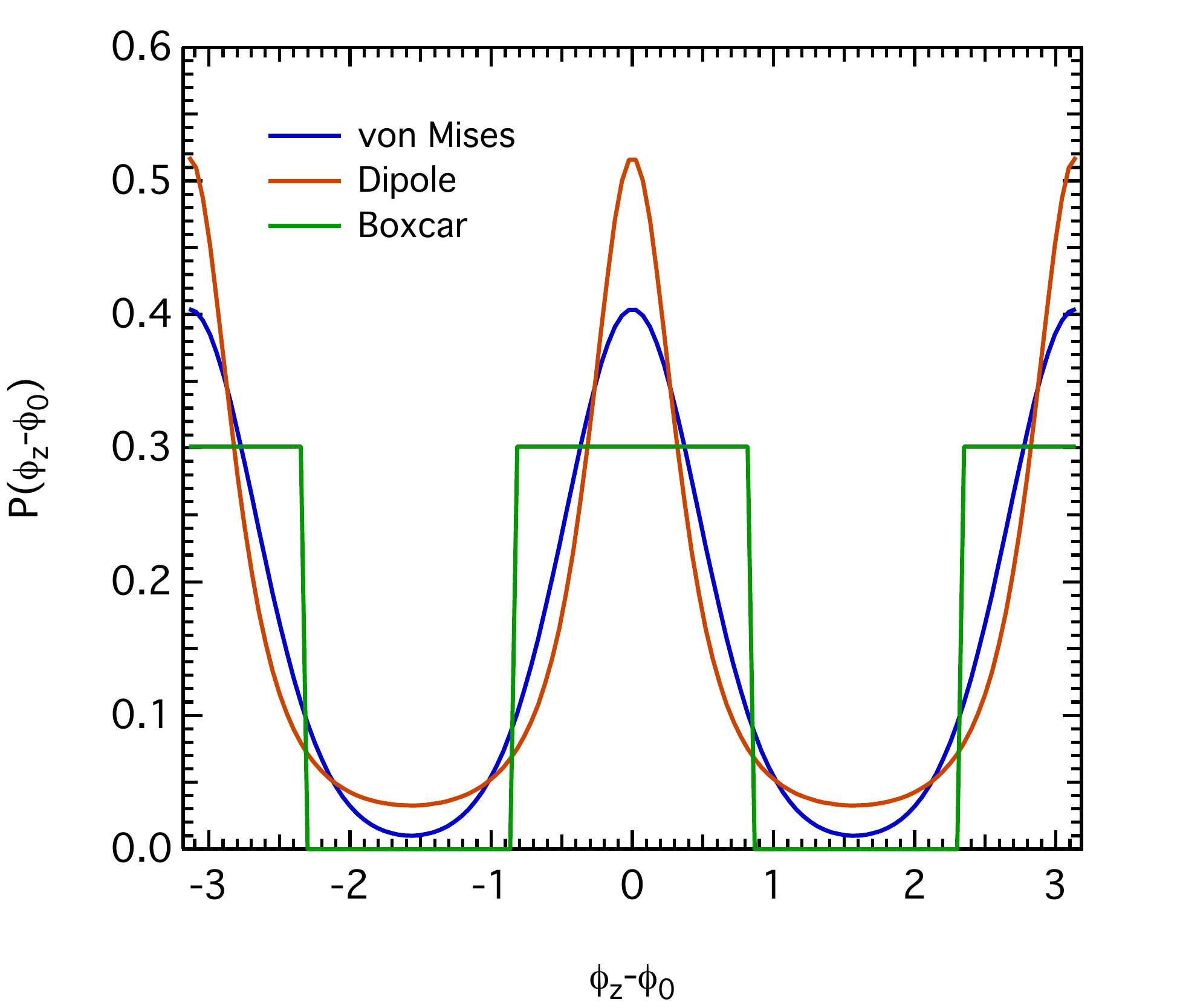}
\caption{The three models for the angular distribution of the power
  spectrum of fluctuations in the scattering screen towards
  \sgra. Model~1 is an amphidirectional von Mises distribution,
  Model~2 is the dipole model of \cite{Johnson2016a}, and Model~3 is a
  periodic boxcar distribution. For each model, the concentration
  parameter is chosen such that the anisotropy of the scattering
  kernel at long wavelengths matches that of \sgra.
  \label{fig:three_models}}
\end{center}
\end{figure}

\subsection{Approximate Expressions}

Using a combination of Taylor expansions and Pad{\'e} approximations, we found a number of approximate relations for the anisotropy parameter $\zeta_0$ and the two coefficients ${\cal B}_{\rm maj}$ and ${\cal B}_{\rm min}$, with an accuracy $\lesssim 1$\% for $k_\zeta\le 10$ and $3/2\le\alpha\le 2$. 

We write the anisotropy parameter $\zeta_0$, for all models, as
\begin{equation}
  \zeta_0=\frac{\sum_{i=0}^{6}\zeta_{{\rm u},i} k_\zeta^i}{\sum_{i=0}^{6}\zeta_{{\rm d},i} k_\zeta^i}\;.
\end{equation}
We write the coefficient ${\cal B_{\rm maj}}$, for all models, as
\begin{equation}
{\cal B}_{\rm maj}=\frac{2^{3-\alpha}}{\alpha \Gamma(1+\alpha/2)}
\left[2\alpha-3-2(\alpha-2)
\frac{\sum_{i=0}^{6}{\cal B}_{{\rm u},i} k\zeta^i}{\sum_{i=0}^{6}{\cal B}_{{\rm d},i} k_\zeta^i}\right]\;,
\end{equation}
Finally, we write a similar expression for the coefficient ${\cal B_{\rm min}}$ for the dipole model, i.e.,
\begin{equation}
{\cal B}_{\rm min}=\frac{2^{3-\alpha}}{\alpha \Gamma(1+\alpha/2)}
\left[2\alpha-3-2(\alpha-2)
\frac{\sum_{i=0}^{6}{\cal M}_{{\rm u},i} k\zeta^i}{\sum_{i=0}^{6}{\cal M}_{{\rm d},i} k_\zeta^i}\right]\;,
\end{equation}
but, for the von Mises and boxcar models, we write
\begin{equation}
{\cal B}_{\rm min}=\frac{2^{3-\alpha}}{\alpha \Gamma(1+\alpha/2)}
\left[\frac{\sum_{i=0}^{6}{\cal M}_{{\rm u},i} k\zeta^i}{\sum_{i=0}^{6}{\cal M}_{{\rm d},i} k_\zeta^i}\right]^{-2(\alpha-2)}\;,
\end{equation}

The coefficients of the Pad{\'e} approximations for the various expressions and for the three models are given in the Appendix. 

\section{Ensemble-Average Visibilities}

With the definitions discussed in the previous two sections, we can
now calculate the effects of interstellar scattering on the images of
scatter-broadened sources and the corresponding interferometric
observables. In this and in the following sections, we will mostly use
the framework and notation of~\cite{Johnson2016a}.

If we denote by $\vec{b}$ the baseline between two stations of an
interferometric array, by $V_{\rm ea}(\vec{b})$ the complex
ensemble-average visibility measured on this baseline, and by $V_{\rm
  src}(\vec{b})$ the complex visibility of the unscattered source,
then~\citep{Coles1987}
\begin{equation}
V_{\rm ea}(\vec{b})=V_{\rm src}(\vec{b})\,\exp\left[-\frac{1}{2}D_{\phi}
   \left(\frac{\vec{b}}{1+M}\right)\right]\;.
   \label{eq:Vea}
\end{equation}
Hereafter, we will use $b=\vert\vec{b}\vert$ to denote the length of the baseline vector. We will consider first the two limiting cases, $b\ll \rin$ and $b\gg\rin$.

\subsection{The $b\ll \rin$ regime}

At long wavelengths (i.e., greater than a few cm in the case of \sgra),
where we expect $b\ll\rin$, the ensemble average
visibility reduces to (see eq.~[\ref{eq:Dsmallr}]) 
\begin{eqnarray}
\lim_{\rdif\ll\rin}V_{\rm ea}(\vec{b})&=&V_{\rm src}(\vec{b})\exp\left[
    -\frac{{\cal C}}{4(1+M)^2}\left(\frac{b}{\rin}\right)^2\right.\nonumber\\
 &&\qquad\qquad   \left.
   \times\{1+\zeta_0\cos[2(\phi-\phi_0)] \}\right]\;.
   \label{eq:Vea_smallrdif}
\end{eqnarray}
Traditionally, the ensemble-average visibility amplitude of a scatter-broadened source is modeled in terms of an anisotropic function of the form~\citep{Narayan1989a,Bower2004}
\begin{equation}
V_{\rm ea}(u',v')=V_{\rm src}(u',v')\exp
\left\{-\frac{\pi^2}{4\ln 2}\left[
\left(u' \thetapar\right)^{\beta-2}+\left(v' \thetaperp\right)^{\beta-2}\right]
\right\}\;,
   \label{eq:VeaGaussrot}
\end{equation}
where $u'$ and $v'$ are baseline lengths in units of the wavelength of
observations in a coordinate system rotated to match the position
angle $\phi_{\rm PA}$ of the scattered image, and $\thetapar$
and $\thetaperp$ are the FWHM angular sizes of the scattered image
along the major and the minor axis of anisotropy, respectively. For
anisotropic scattering, we also define the degree of anisotropy of the
scattering kernel as the ratio of the angular sizes along the major
and minor axis of the kernel, i.e.,
\begin{equation}
A\equiv\frac{\thetapar}{\thetaperp}\;.
\label{eq:anisotropy}
\end{equation}

The case considered in this subsection, $b\ll \rin$, corresponds to
$\beta=4$ (structure function varying quadratically with
distance). Rotating the coordinate system to the usual $(u,v)$ plane,
with the position angle $\phi_0$ measured as an angle North of East,
and transforming to polar coordinates, $(u,v)
\equiv(b/\lambda)\,(\cos\phi,\sin\phi)$, we can write
expression~(\ref{eq:VeaGaussrot}) as
\begin{eqnarray}
V_{\rm ea}(\vec{b})&=&V_{\rm src}(\vec{b})\exp
\left[-\frac{\pi^2 b^2\thetapar^2}{\lambda^2(8\ln 2)}\left(\frac{A^2+1}{A^2}\right)
\right.\nonumber\\
&&\qquad\qquad\left.
\left(1+\frac{A^2-1}{A^2+1}\cos[2(\phi-\phi_0)]\right) \right]\;.
\end{eqnarray}
In order to put this expression in the same form as equation~(\ref{eq:Vea_smallrdif}), we  identify
\begin{equation}
  \lim_{b\ll\rin} A=\left(\frac{1+\zeta_0}{1-\zeta_0}\right)^{1/2}\;,
\label{eq:A_smallrdif}
\end{equation}
so that
\begin{equation}
V_{\rm ea}(\vec{b})=V_{\rm src}(\vec{b})\exp
\left[-\frac{\pi^2 b^2 \thetapar^2}{\lambda^2(4\ln 2)}
\left(\frac{1+\zeta_0\cos[2(\phi-\phi_0)]}{1+\zeta_0}\right) \right]\;.
\label{eq:VeaGauss}
\end{equation}

Comparing equations~(\ref{eq:Vea_smallrdif}) and (\ref{eq:VeaGauss}), we find that the angular size of a scatter-broadened point source along the major axis of anisotropy, at some wavelength $\lambda_0$ that satisfies the limit $b\ll\rin$, is
\begin{eqnarray}
  \thetaparzero&=&\sqrt{4\ln 2(1+\zeta_0){\cal C}}\frac{\lambdabar_0}{(1+M)\rin}
  \nonumber\\
  &=&\left[\frac{(1+\zeta_0)\bar{Q}_z \Gamma(1-\alpha/2)\ln 2}{2}\right]^{1/2}
  \frac{\lambdabar_0^2}{(1+M)\pi \rin^2}\;,
\label{eq:theta_scat}
\end{eqnarray}
where, in the last expression, we used relation~(\ref{eq:calC}). We
can solve equation~(\ref{eq:theta_scat}), using also the
definitions~(\ref{eq:anisotropy}) and (\ref{eq:A_smallrdif}), to
express the amplitude of the power spectrum of turbulence in terms of
the angular sizes of the scattering kernel at some reference
wavelength $\lambdabar_0$ as
\begin{equation}
\bar{Q}_z=\frac{(1+M)^2 \pi^2
    (\thetaparzero^2+\thetaperpzero^2)}
       {\Gamma(1-\alpha/2)\ln 2}
       \left(\frac{\rin}{\lambdabar_0}\right)^4\;.
\end{equation}

For the scattering screen towards \sgra, in this limit, $A\simeq
2$~\citep[see, e.g.,][]{Bower2004, Bower2006}. Therefore, $\zeta_{\rm
  SGRA,0}\simeq 3/5$.  Furthermore, connecting
equation~(\ref{eq:theta_scat}) to equation~(\ref{eq:imageSGRA}) and
setting $M=(2.7~$kpc$)/(5.6~$kpc$)\simeq 0.5$, as inferred from
observations of the Galactic Center magnetar~\citep{Bower2014a}, we obtain
\begin{equation}
\bar{Q}_z=2.5\times 10^{-12}\left[\Gamma\left(1-\frac{\alpha}{2}\right)\right]^{-1} \left(\frac{\rin}{\mbox{cm}}\right)^4\;.
\end{equation}
Note that, in principle, long wavelength observations of the ensemble
average visibilities allow us only to infer the ratio
$\bar{Q}_z/\rin^4$. Extrapolating the model to smaller wavelenghts for
which the diffractive scale is compareble to or smaller than the inner
scale of turbulence requires an independent measurement of the size of
the inner scale.

\subsection{The $b\gg \rin$ regime}

In this regime, the ensemble average visibility cannot be put in the
form~(\ref{eq:VeaGaussrot}), even if we set $\beta=\alpha+2$, which is
the fitting formula often used. Nevertheless, we can use the general
form of the phase structure function, i.e.,
equation~(\ref{eq:larger}), to reach two important conclusions.

First, the orientation of the major axis of the scattering kernel
(defined by the angle $\phi_0$) does not depend on wavelength. In
other words, the position angle for the scattering kernel of
\sgra\ measured at long wavelengths can be used even when
extrapolating the properties of the scattering screen to the
mm-wavelengths of the EHT.

Second, the degree of anisotropy of the scattering kernel evolves with
wavelength.  It proved impossible to convert the general expression of
the phase structure function in this case into a corresponding image
and calculate its FWHM along the two axes of the scattering
kernel. Instead, in order to get an understanding of the dependence of
the characteristic size and anisotropy of the scatter-broadened image
on wavelength, we first use equation~(\ref{eq:Vea}) with the phase
structure function along the major and minor axes of the scatter
broadened image (eqs.~[\ref{eq:Dmaj_larger}] and
[\ref{eq:Dmin_larger}]) to find the baseline lengths $b_{\rm maj}$ and
$b_{\rm min}$ for which the exponent in the right-hand side of
equation~(\ref{eq:Vea}) becomes equal to $-(1/2)$, i.e.,
\begin{equation}
  b_{\rm maj}=(1+M)\rin
  \left[\frac{2}{(1+\zeta_0){\cal B}_{\rm maj}{\cal C}}\right]^{1/\alpha}
  \label{eq:bpar}
\end{equation}
and
\begin{equation}
    b_{\rm min}=(1+M)\rin
    \left[\frac{2}{(1-\zeta_0){\cal B}_{\rm min}{\cal C}}\right]^{1/\alpha}
    \label{eq:bperp}
\end{equation}

These, of course, correspond to the baseline lengths that resolve the
diffractive scale of the scattering screen along the major and minor axes of the scatter broadened image. We then use these lengths to
define characteristic image sizes via
\begin{eqnarray}
  \thetapar&\equiv& \sqrt{8\ln2}\frac{\lambdabar}{b_{\rm maj}}=
  \sqrt{8\ln2}\frac{\lambdabar}{(1+M)\rin}
    \left[\frac{(1+\zeta_0){\cal B}_{\rm maj}{\cal C}}{2}\right]^{1/\alpha}
\label{eq:theta_parallel}
\end{eqnarray}
and
\begin{eqnarray}
  \thetaperp&\equiv& \sqrt{8\ln2}\frac{\lambdabar}{b_{\rm min}}=
  \sqrt{8\ln2}\frac{\lambdabar}{(1+M)\rin}
  \left[\frac{(1-\zeta_0){\cal B}_{\rm min}{\cal C}}{2}\right]^{1/\alpha}
\label{eq:theta_perp}
\end{eqnarray}
Combining these two equations, we can calculate the degree of anisotropy
in the limit $b\gg \rin$
\begin{equation}
\lim_{b\gg \rin} A=\left[\frac{(1+\zeta_0){\cal B}_{\rm
      maj}}{(1-\zeta_0){\cal B}_{\rm min}}\right]^{1/\alpha}\;.
\end{equation}
This is different than the $b\ll\rin$ regime both because of the
exponent being $1/\alpha$ as opposed to $1/2$ and because, in general,
${\cal B}_{\rm maj}/{\cal B}_{\rm min}\ne 1$.

\subsection{The General Case}

In the general case, we write the ensemble average
visibility~(\ref{eq:Vea}) using the approximate
expression~(\ref{eq:D_general}) for the phase structure function. We
then follow the same procedure as in equations~(\ref{eq:bpar}) and
(\ref{eq:bperp}) to find the baseline lengths that resolve the diffractive
scale along the major and minor axes of anisotropy as
\begin{eqnarray}
  b_{\rm maj}&=&(1+M)\rin
  \left(\frac{2}{\alpha {\cal B}_{\rm maj}}\right)^{1/(\alpha-2)}
  \nonumber\\
  &&\qquad \left\{ \left[1+4^{1/(2-\alpha)}
    \frac{{\cal B}_{\rm maj}^{2/(\alpha-2)} \alpha^{\alpha/(\alpha-2)}}
         {{\cal C}(1+\zeta_0)}\right]^{2/\alpha}-1\right\}^{1/2}
  \label{eq:bpar_general}
\end{eqnarray}
and
\begin{eqnarray}
  b_{\rm min}&=&(1+M)\rin
  \left(\frac{2}{\alpha {\cal B}_{\rm min}}\right)^{1/(\alpha-2)}
  \nonumber\\
  &&\qquad \left\{ \left[1+4^{1/(2-\alpha)}
    \frac{{\cal B}_{\rm min}^{2/(\alpha-2)} \alpha^{\alpha/(\alpha-2)}}
         {{\cal C}(1-\zeta_0)}\right]^{2/\alpha}-1\right\}^{1/2}\;,
  \label{eq:bperp_general}
\end{eqnarray}
the corresponding image sizes as
\begin{eqnarray}
  \thetapar&\equiv& \sqrt{8\ln2}\frac{\lambdabar}{b_{\rm maj}}
\end{eqnarray}
and
\begin{eqnarray}
  \thetaperp&\equiv& \sqrt{8\ln2}\frac{\lambdabar}{b_{\rm min}}
\end{eqnarray}
and the degree of anisotropy using equation~(\ref{eq:anisotropy}).

\begin{figure}[t]
\begin{center}
\includegraphics[height=7cm,width=8cm]{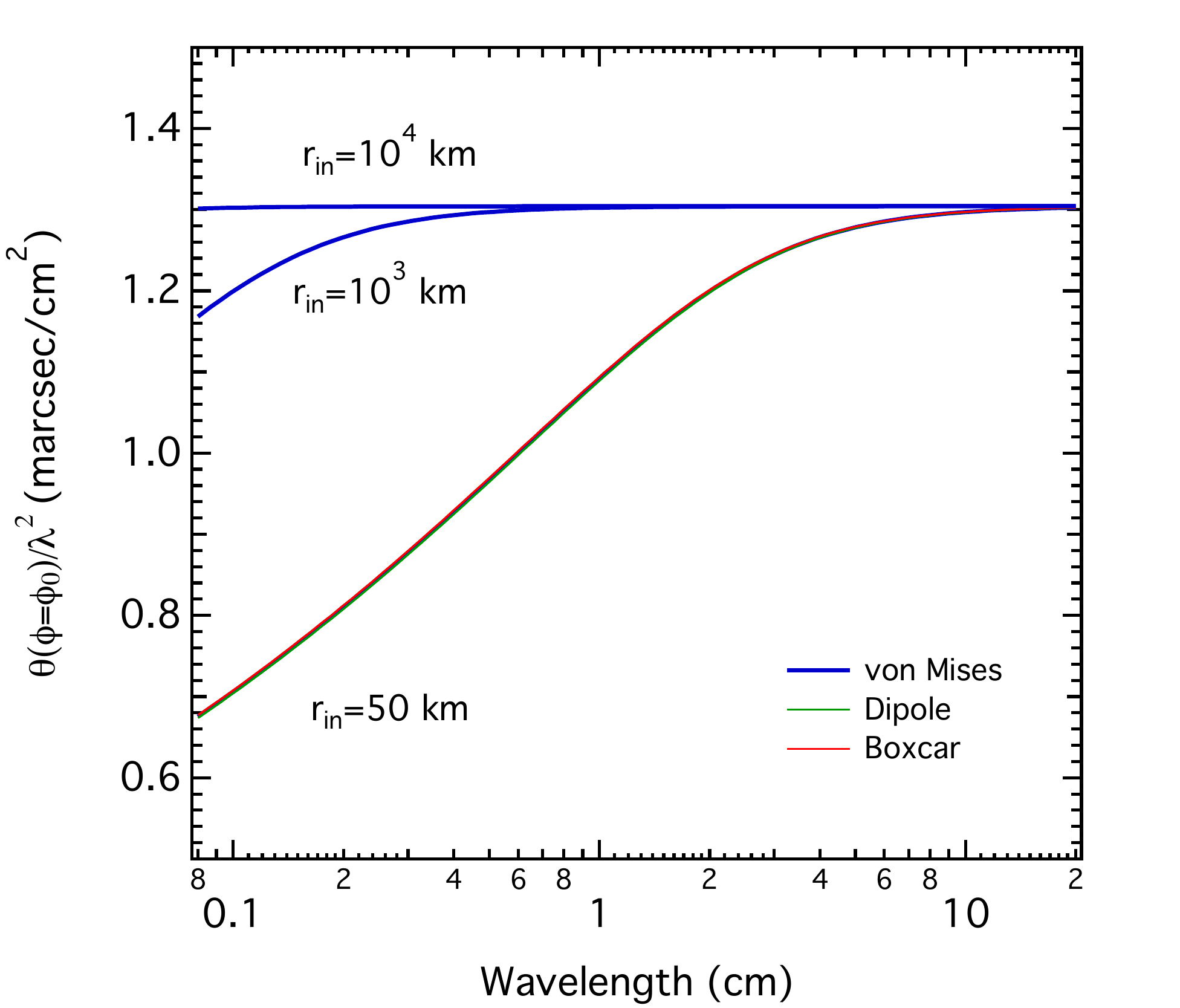}
\includegraphics[height=7cm,width=8cm]{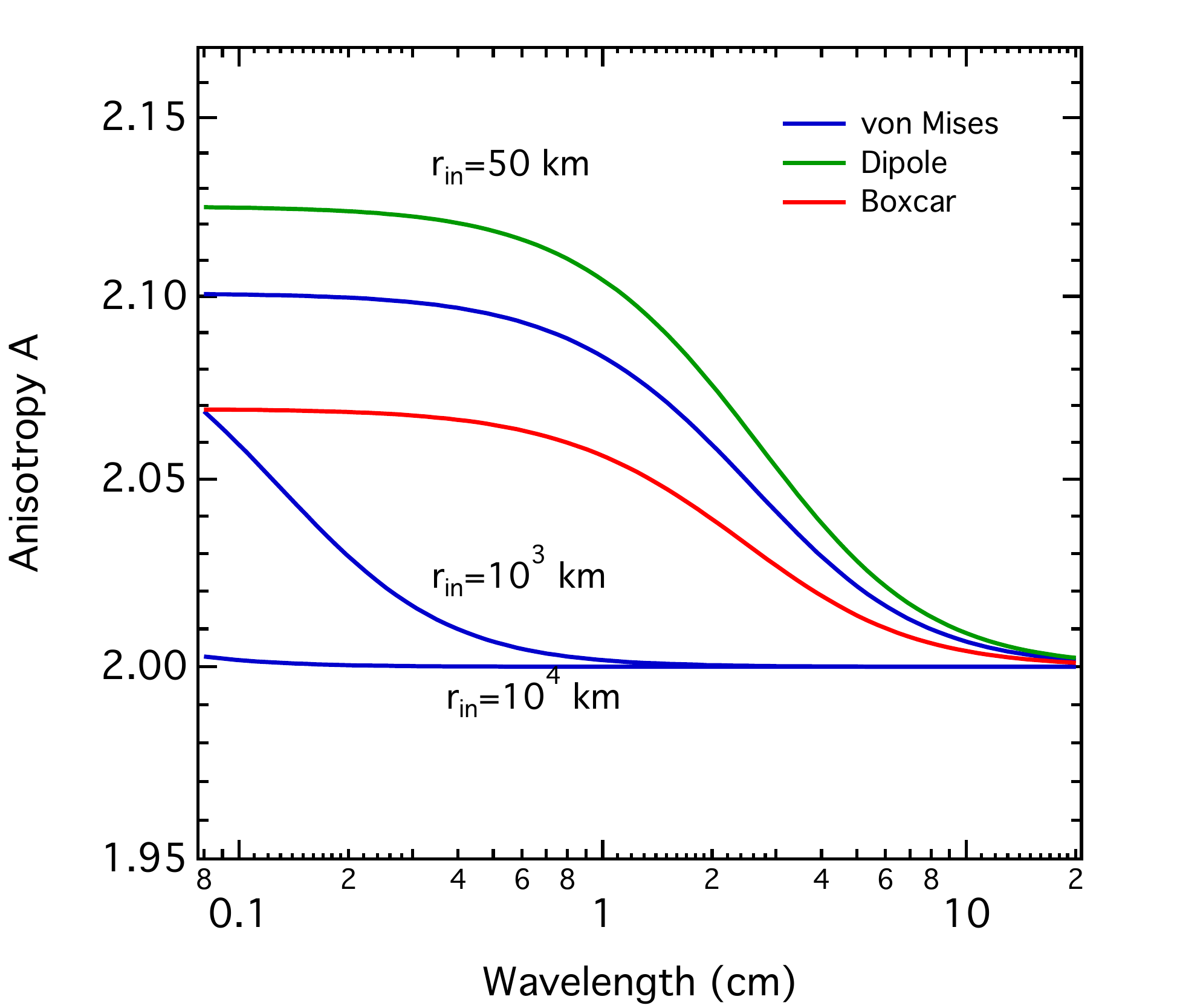}
\caption{{\em (Top)\/} The characteristic size of the scattering kernel for \sgra, along its major axis, divided by the square of the wavelength, as a function of wavelength. {\em (Bottom)\/} The degree of anisotropy of the scattering kernel,  defined as the ratio of its major to minor axis. The power spectrum of fluctuations is assumed to be Kolmogorov ($\alpha=5/3$), with the remaining parameters of the screen inferred from long-wavelength observations (see \S5.1). The blue curves correspond to three different choices for the inner scale of turbulence $\rin$ in the scattering screen. Depending on $\rin$, the scattering kernel at short wavelengths becomes smaller and more anisotropic compared to what a simple extrapolation from long wavelengths predicts. The different colors correspond to the three models for the wandering of the direction of anisotropy for $\rin=50$~km. Even though the major axis of the scattering kernel depends very weakly on the angular model, the minor axis and, hence, the anisotropy, does depend at the $\sim 5$\% level.
  \label{fig:kernel_size}}
\end{center}
\end{figure}

\begin{figure*}[t]
\begin{center}
\includegraphics[height=7cm,width=8cm]{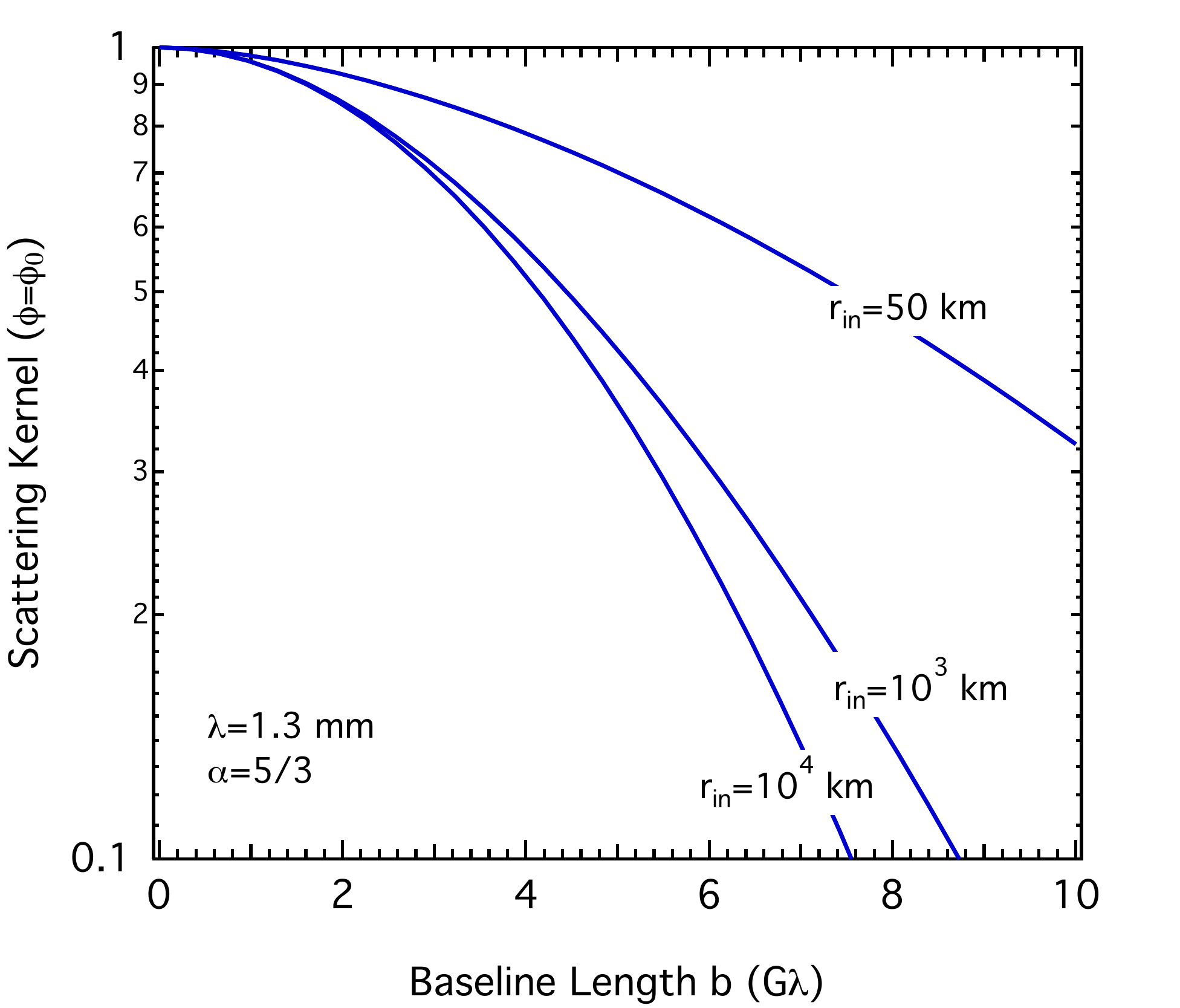}
\includegraphics[height=7cm,width=8cm]{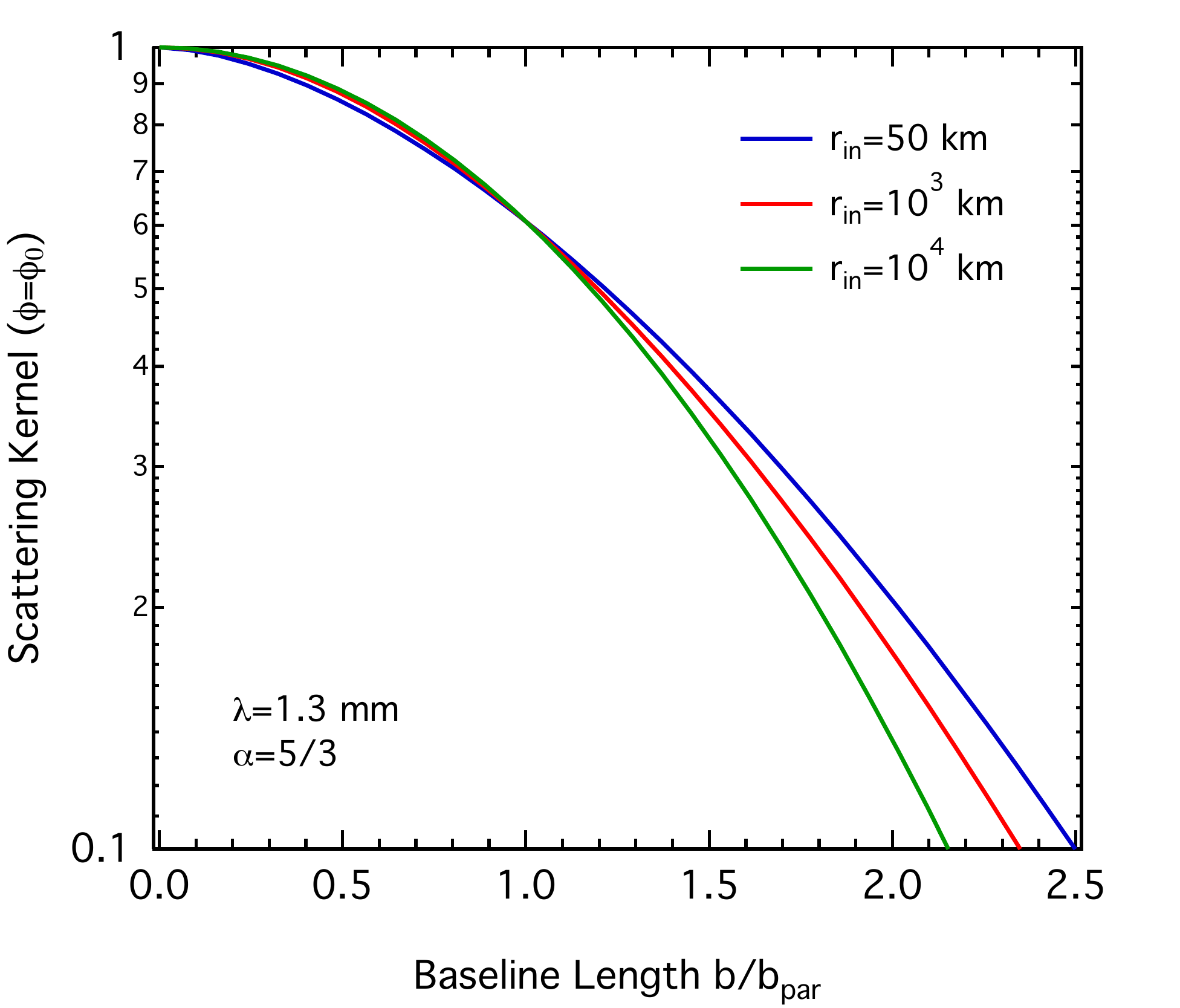}
\caption{The scattering kernel of \sgra, along its major axis, at
  1.3~mm. All other parameters are the same as in
  Figure~\ref{fig:kernel_size}. In the left panel, the kernel is
  plotted as a function of baseline length in units of G$\lambda$; in
  the right panel, the baseline lengths have been normalized to the
  characteristic length $b_{\rm maj}$ given by
  equation~(\ref{eq:bpar_general}). A smaller inner scale of turbulence
  leads to reduced broadening, and also alters the functional form of
  the scattering kernel.
  \label{fig:kernel_sgra}}
\end{center}
\end{figure*}

Figure~\ref{fig:kernel_size} shows the wavelength dependence of the
major axis of the scattering kernel for \sgra\ as well as its
anisotropy for a Kolmogorov spectrum of fluctuations ($\alpha=5/3$),
for the three models of magnetic-field wandering, and for the parameters we obtained in
\S4.1. Depending on the (unknown) inner scale of turbulence in the
scattering screen, the size of the scattering kernel at the EHT
wavelengths may be reduced by as much as 50\% from the simple
extrapolation of the $\lambda^2$ dependence found at long
wavelengths. Also, the degree of anisotropy of the scattering kernel
increases with decreasing wavelength, depending again on the inner
scale of turbulence. Even though the major axis of the scattering kernel depends very weakly on the angular model, the minor axis and, hence, the anisotropy, does depend weakly on them at the $\sim 5$\% level.

Figure~\ref{fig:kernel_sgra} shows the dependence of the scattering
kernel on baseline length at a wavelength of 1.3~mm; all other
parameters are the same as in Figure~\ref{fig:kernel_size}. As
expected from the above discussion, the size of the scattering kernel,
which determines the baseline at which the visibility cuts off, varies
with the choice of $\rin$. Moreover, the functional form of the
scattering kernel evolves with wavelength, becoming shallower (in
$u-v$ space) once the diffractive length becomes larger than the inner
scale of turbulence. The largest EHT baselines are $\simeq 6$G$\lambda$ and $\simeq 9$G$\lambda$ in the E-W and N-S directions, respectively. As a result, the effects of a finite inner scale of turbulence may significantly affect the interpretation and deblurring of EHT observations of \sgra.

\section{Refractive Effects}

In the average image regime, which is relevant for single-epoch VLBI observations of \sgra, the images and interferometric observables will exhibit fluctuations due to refractive effects~\citep{Narayan1989a,Goodman1989,Gwinn2014,Johnson2015,Johnson2016a}. These effects are dominated by density fluctuations on scales comparable to the refractive scale $\simeq r_{\rm ref}$. Because of the large separation between the diffractive and refractive scales (see Fig.~\ref{fig:scater_scales}), one can estimate the refractive effects using a simplified formalism for scattering developed originally for pulsars by \cite{Blandford1985} and applied recently to extended sources and interferometry by \cite{Johnson2016a}. 

In this framework, the variance of the complex visibility on baseline $\vec{b}$ caused by refractive substructure is estimated by the following expression \citep[see eq.~16 of][]{Johnson2016a},
\begin{equation}
\sigma_{\rm ref}^2 \approx \frac{\lambdabar^2 \rF^4}{4\pi^2}\int d^2\vec{q}
\left[\vec{q}\cdot\left(\vec{q}+\frac{\vec{b}}{D\lambdabar}\right)\right]^2
\left\vert V_{\rm ea}\left(D\lambdabar \vec{q}+\vec{b}\right)\right\vert^2 Q(\vec{q})\;.
\label{eq:sref2}
\end{equation}
For comparison with VLBI observations, this expression must be
modified to remove the variance that arises from modulation of the
total image flux density and image wander (caused, respectively, by
refractive focusing/defocusing and an overall ray deflection;
\citealt{Blandford1985}). These effects are degenerate with standard
VLBI calibration (in the absence of absolute flux calibration and
absolute phase referencing, which are challenging at high
frequencies). Thus, we will also define the \emph{renormalized
  refractive noise}, which removes these two contributions. For
example, the renormalized refractive noise on a zero baseline is zero
because all variations on this baseline reflect modulation of the
total flux density. Differences between the total refractive noise and
the renormalized refractive noise are significant for baselines that
do not heavily resolve the source.

Because of the complexity of the integrals in
equation~(\ref{eq:sref2}), we evaluate them numerically, using the
approximate expression~(\ref{eq:D_general}) for the ensemble
averaged visibility, but the complete expressions for the power
spectra.  The refractive variance on a given baseline depends on three
parameters: the power-law index of turbulence $\alpha$, the ratio of
the characteristic baseline length $b_{\rm maj}$ (see
eq.[\ref{eq:bpar_general}]) that resolves the image along the
direction of anisotropy ($\phi=\phi_0$) to the inner scale of
turbulence,
\begin{equation}
\mu_{\rm d}\equiv \frac{b_{\rm maj}}{\rin}\;,
\end{equation}
and the ratio
\begin{equation}
\rho_0\equiv \frac{D\lambdabar}{\rin^2} = (1+M)
\left(\frac{r_{\rm F}}{\rin}\right)^2\;,
\end{equation}
which measures the number of circular patches of size equal to $\rin$ that fits within a Fresnel circle. For \sgra, at $\lambda=1$~cm,
\begin{equation}
\rho_0 \simeq 1.3\times 10^{5}(1+M)
\left(\frac{D}{2.7~\mbox{kpc}}\right)
\left(\frac{\lambda}{1~\mbox{cm}}\right)
\left(\frac{\rin}{10^3~\mbox{km}}\right)^{-2}.
\end{equation}

Figure~\ref{fig:model1_refnoise} shows the dependence of the rms
refractive visibility on baseline length for a point source, and for
parameters that are expected for \sgra. At small baseline lengths, the
major cause of rms visibility fluctuations is flux variations and
image wandering, neither of which is observable by a mm-VLBI
experiment such as the EHT. On the other hand, at baseline lengths
larger than the diffractive scale, the rms visibility is primarily
caused by the refractive substructure in the image introduced by
scattering and attains a characteristic power-law dependence on
baseline length. In general, the rms refractive visibility at large
baselines is higher along the major axis of anisotropy, since there is
more power of density fluctuations along that orientation.

In Figure~\ref{fig:model1_refnoise}, the baseline length is normalized
to the characteristic baseline length $b_{\rm maj}$ (see
eq.[\ref{eq:bpar_general}]) that resolves the image along the
direction of anisotropy ($\phi=\phi_0$). For this reason, the
transition to the power-law dependence occurs at longer baseline
lengths for orientations that are perpendicular to the direction of
anisotropy ($\phi=\phi_0+\pi/2$), since the corresponding diffractive
scale is larger.

\begin{figure}[t]
\begin{center}
\includegraphics[height=7cm,width=8cm]{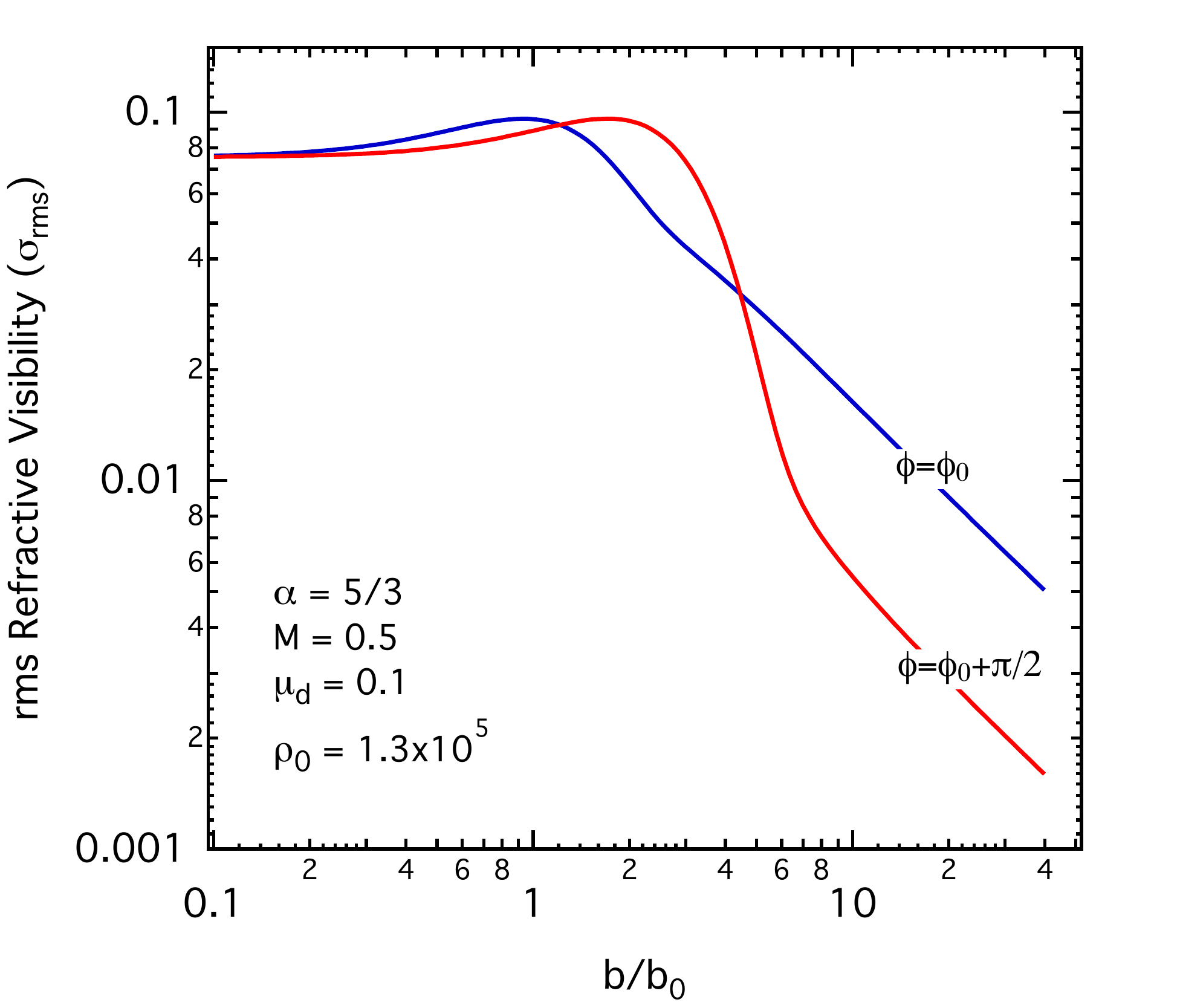}
\caption{The rms refractive visibility ($\sigma_{\rm ref}$) as a function of baseline length for parameters of the scattering screen that are typical for \sgra\ at mm-wavelengths. The blue and red curves show the result for orientations that are parallel and perpendicular to axis of anisotropy. In both cases, the baseline length is normalized to the characteristic baseline length $b_{\rm maj}$ (see eq.[\ref{eq:bpar_general}]) that resolves the image along the direction of anisotropy (i.e., at $\phi=\phi_0$). At large baselines, the rms refractive visibility is higher along the direction of anisotropy, which corresponds (by definition) to a larger amplitude of density fluctuations in the scattering screen. }
  \label{fig:model1_refnoise}
\end{center}
\end{figure}

\begin{figure}[t]
\begin{center}
\includegraphics[height=7cm,width=8cm]{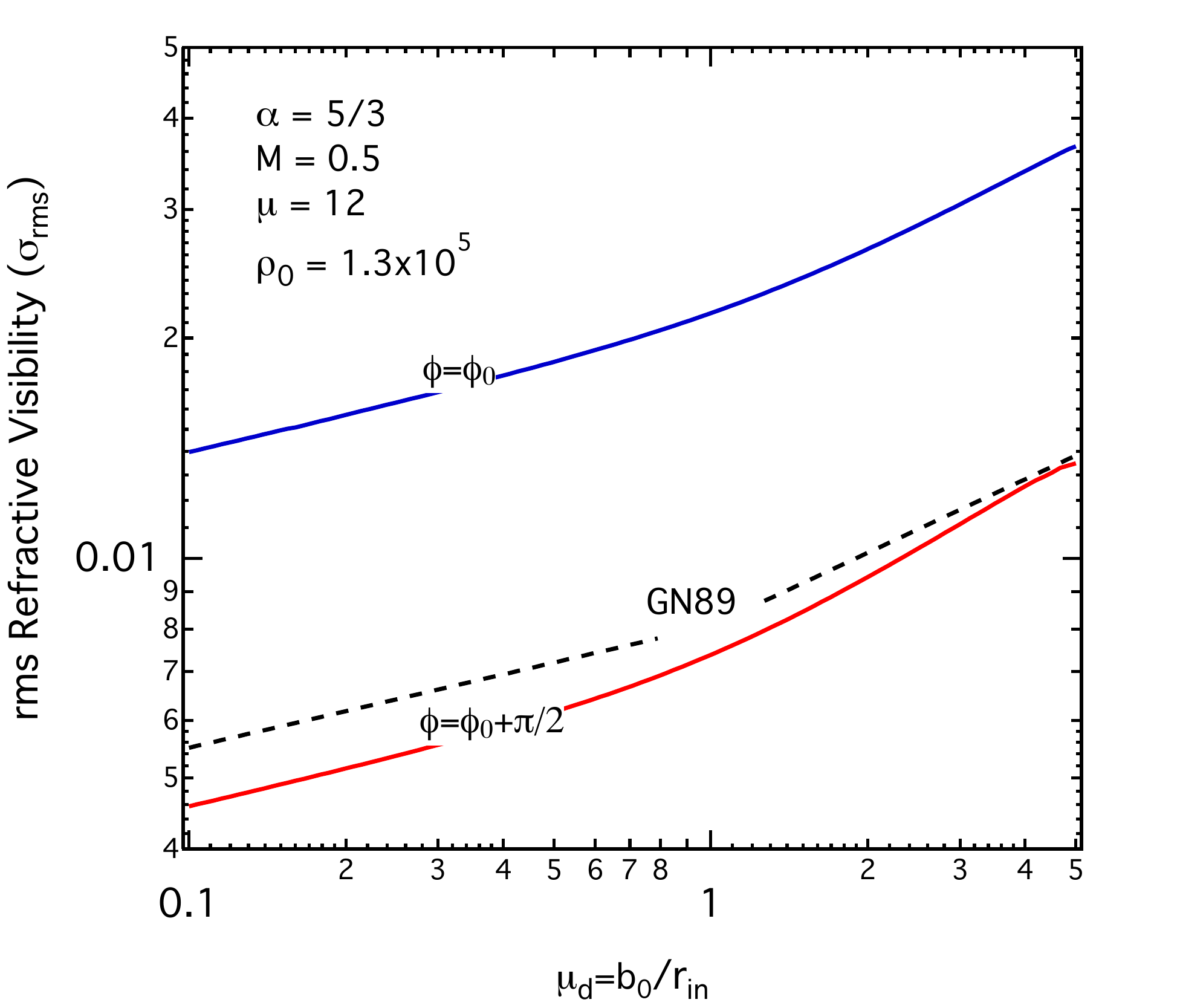}
\caption{The rms refractive visibility ($\sigma_{\rm ref}$) evaluated at $b=12b_{\rm maj}$ as a function of the ratio of the characteristic baseline length $b_{\rm maj}$ to the inner scale of turbulence. The parameters of the scattering screen are the same as in Fig~\ref{fig:model1_refnoise}. The blue and red curves show the result for orientationw that are parallel and perpendicular to the axis of anisotropy. The dotted lines show the expressions derived by \cite{Goodman1989} for anisotropic scattering, in the two different regimes $\mu_{\rm d}\ll 1$ and $\mu_{\rm d}\gg 1$.
  \label{fig:model1_mud}}
\end{center}
\end{figure}

\begin{figure}[t]
\begin{center}
\includegraphics[height=7cm,width=8cm]{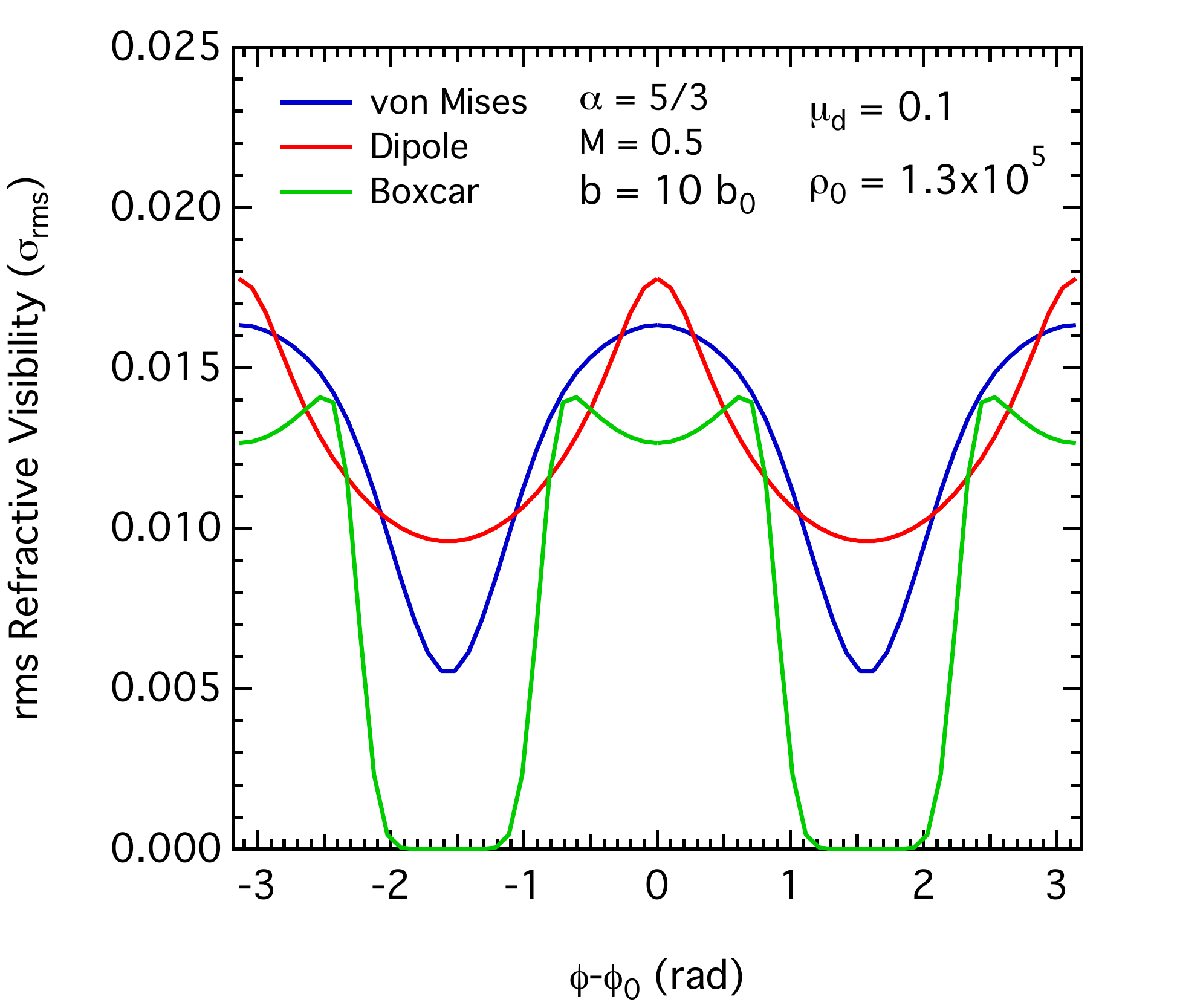}
\caption{The rms refractive visibility ($\sigma_{\rm ref}$) evaluated at $b=10 b_{\rm maj}$ as a function of the orientation angle with respect to the major axis of the scattering kernel, for the three different models of the angular dependence of the power spectrum of density fluctuations. The parameters of the scattering screen are the same as in Fig~\ref{fig:model1_refnoise}. The rms of the refractive visibility depends very strongly on the details of the angular dependence of the power spectrum of fluctuations.
  \label{fig:models_angle}}
\end{center}
\end{figure}

Figure~\ref{fig:model1_mud} shows the dependence of the rms refractive visibility for a long baseline ($b=12 b_{\rm maj}$) on the ratio $\mu_{\rm d}$ of the characteristic baseline length $b_{\rm maj}$ to the inner scale of turbulence.  It also compares the numerical results to the analytical expressions derived by \cite{Goodman1989} for the case of isotropic scattering in two regimes, depending on whether the diffractive scale is larger or smaller than the inner scale of turbulence ($\mu_{\rm d}\ll 1$ or 
$\mu_{\rm d}\gg 1$).  In our notation, these two expressions become for  $\mu_{\rm d}\ll 1$,
\begin{equation}
\sigma_{\rm ref}=\left[2^{\alpha-2}
\frac{\Gamma(1+\alpha/2)}{\Gamma(1-\alpha/2)}\right]^{1/2}
\left(\frac{\rho_0}{1+M}\right)^{\alpha/2-1}
\mu^{-\alpha/2}
\mu_{\rm d}^{1-\alpha/2},
\end{equation}
and for  $\mu_{\rm d}\gg 1$,
\begin{equation}
\sigma_{\rm ref}=\left[2^{\alpha-2}
\frac{\Gamma(1+\alpha/2)\Gamma(4/\alpha)}{\Gamma(1-\alpha/2)}\right]^{1/2}
\left(\frac{\rho_0}{1+M}\right)^{\alpha/2-1}
\mu^{-\alpha/2}
\mu_{\rm d}^{2-\alpha};,
\end{equation}
where $\mu\equiv b/b_{\rm maj}$. As Figure~\ref{fig:model1_mud} shows, the dependence of the rms refractive visibility on the ratio $\mu_{\rm d}=b_{\rm maj}/r_{\rm in}$  
follows, in all orientations, the general trend exhibited in the case of isotropic scattering. However, as expected, the rms refractive visibility is, in general, larger parallel to the major scattering axis and smaller perpendicular to it than what the above expressions predict. This is a direct consequence of the angular dependence of the power spectrum of density fluctuations, which determines both the anisotropy in the scattering kernel and the amplitude of refractive fluctuations.

We now turn to an important feature of refractive effects. To recall, equation~(\ref{eq:D_general}) shows that the scattering kernel that figures in the ensemble average visibilities depends primarily on the overall image anisotropy parameter $A$; it is insensitive to the detailed angular dependence of the power spectrum of density fluctuations, $P(\phi_z,\phi_0)$. By contrast, the rms fluctuations in refractive visibilities depend very strongly on $P(\phi_z,\phi_0)$. Figure~\ref{fig:models_angle} shows the angular dependence of the rms refractive visibility for parameters typical of the scattering screen towards \sgra, for the three different models of the angular power spectrum. Clearly, the model of the power spectrum with the box-car angular distribution  results in zero refractive visibilities at angles perpendicular to the major axis of the scattering kernel. We show, in Figure~\ref{fig:images_rin_Pphi}, that the refractive noise does not actually fall to zero because of higher-order terms in the scattering approximation, although the noise is still markedly reduced relative to the von Mises and dipole models (see Figure~\ref{fig:refractive_noise_comparison}).

\begin{figure*}
\begin{center}
\includegraphics[width=0.45\textwidth]{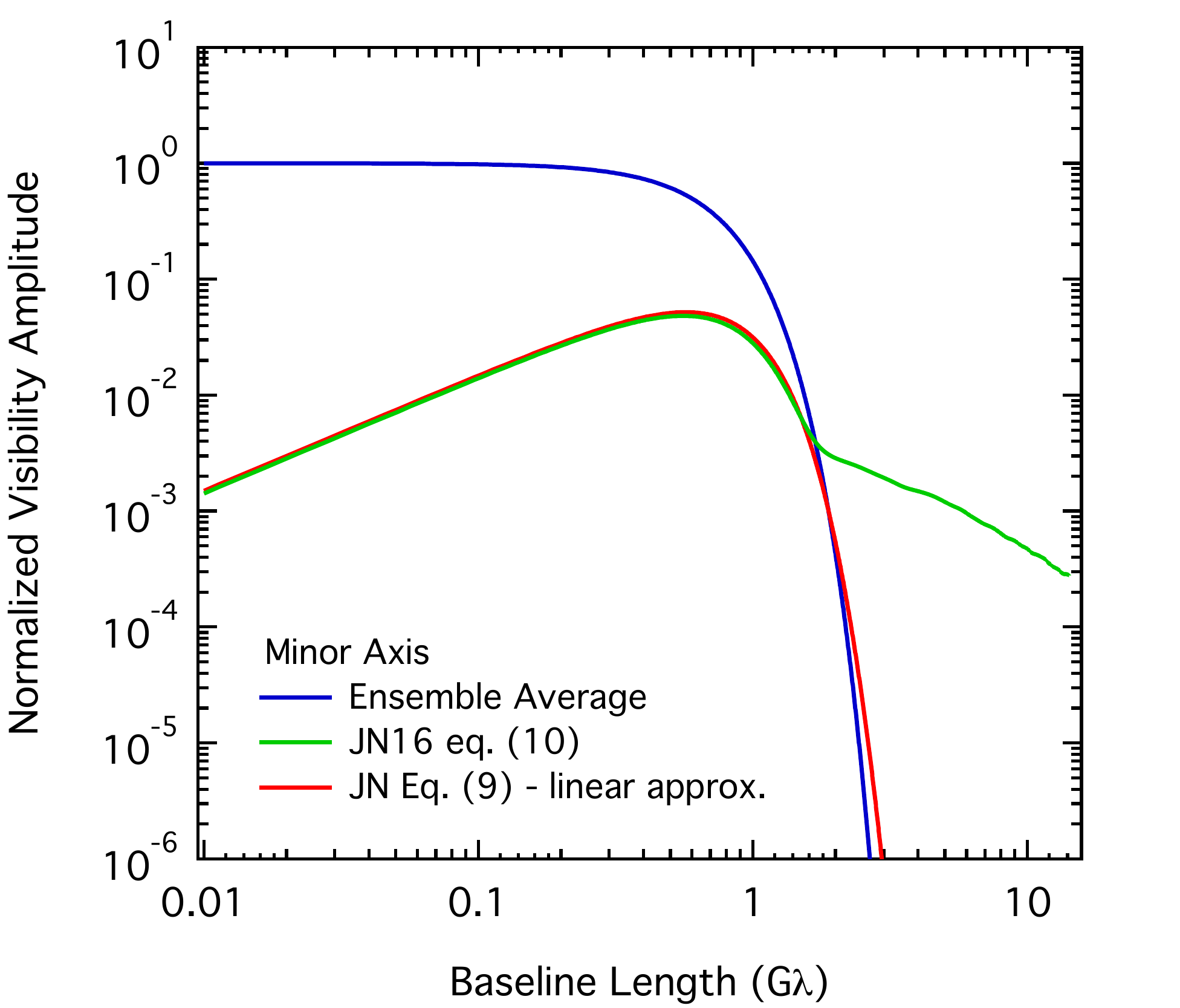}
\includegraphics[width=0.45\textwidth]{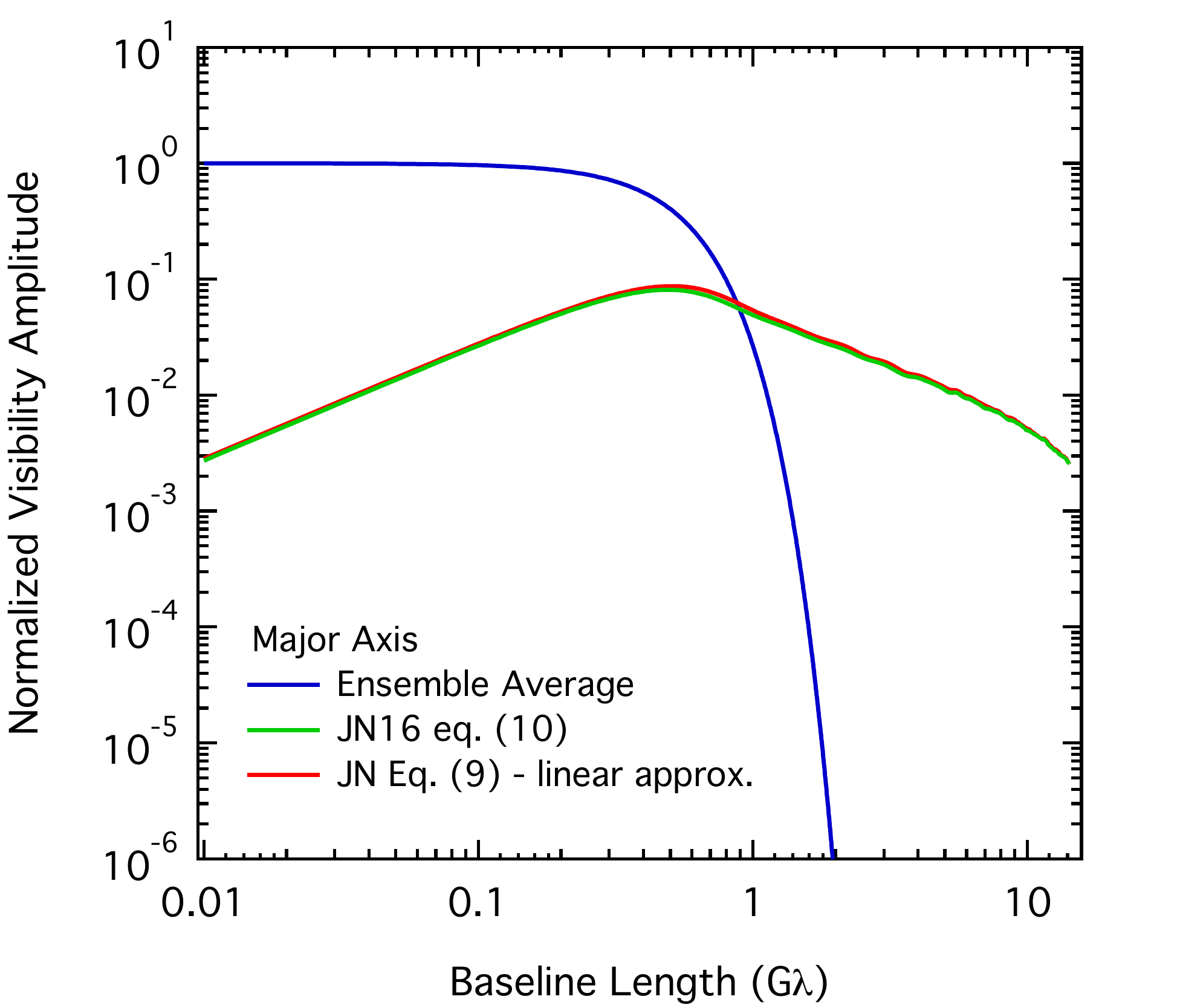}
\caption{Ensemble-average visibilities and the renormalized refractive noise for \sgra\ at 3\,mm with the boxcar model for the magnetic field wandering. Left/right panel shows dependence on baseline length for baselines oriented along the minor/major axis. The plotted curves were calculated from numerical simulations of the scattering (as in later Figures); the refractive noise was estimated using 1000 independent realizations of the scattering. Using the linear approximation \citep[Eq.~10 of][]{Johnson2016a} for the scattering simulations results in almost no power for long baselines along the minor axis, as predicted by Eq.~\ref{eq:sref2} (which also uses the linear approximation). However, including the full expression for refractive steering in the simulations \citep[Eq.~9 of][]{Johnson2016a} produces a small amount of additional refractive noise along this axis, demonstrating the breakdown of the linear approximation. 
}
  \label{fig:images_rin_Pphi}
\end{center}
\end{figure*}

\begin{figure*}
\begin{center}
\includegraphics[height=17cm,width=18cm]{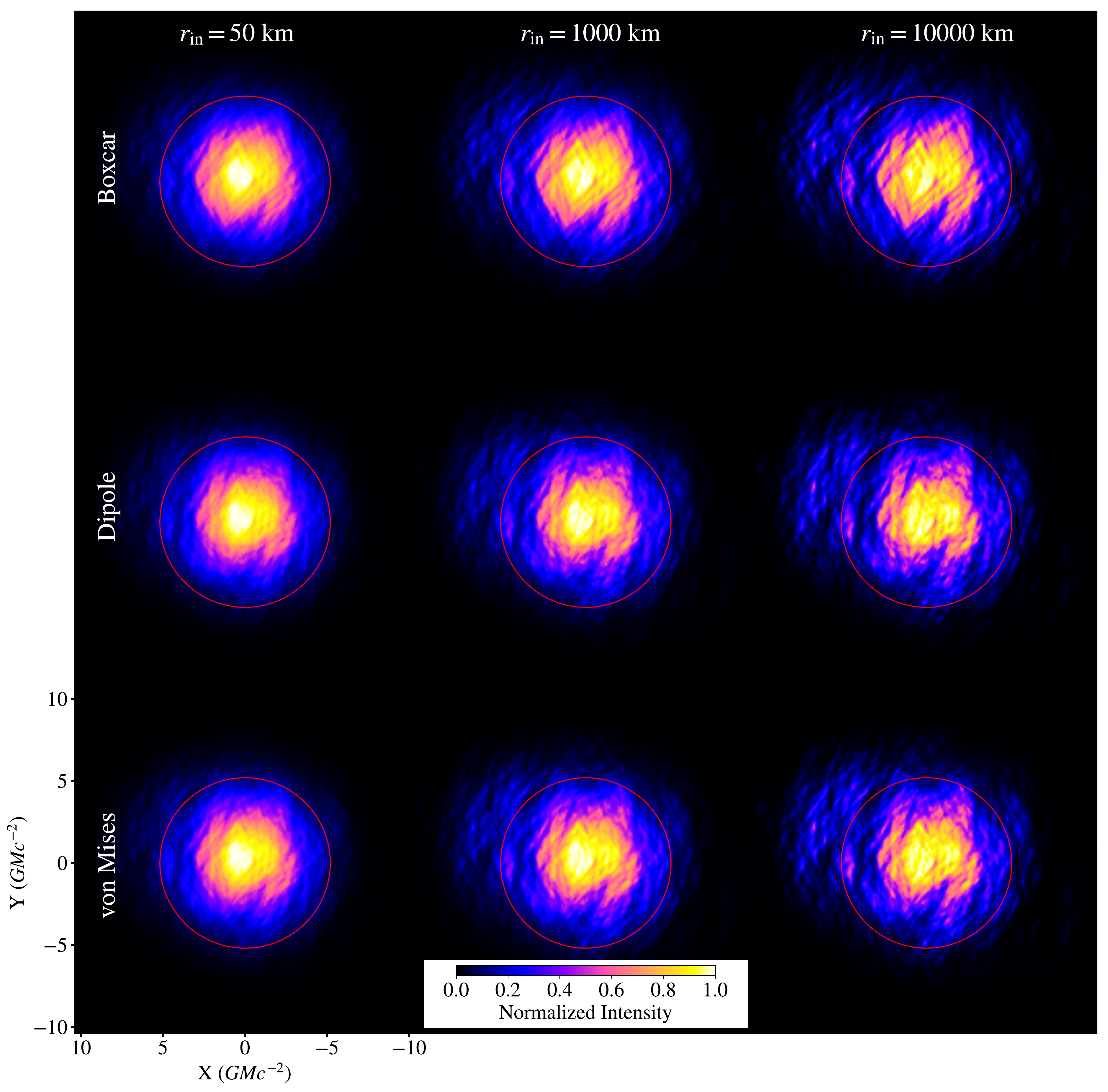}
\caption{Simulated scattered images of \sgra\ at 1.3\,mm. Each image shows a single instantaneous image (i.e., a representative image in the ``Average'' regime). Columns show representative values of the inner scale (50~km, 1{,}000~km, and 10{,}000~km); rows show our three example scattering models. In all cases, the unscattered image is a circular Gaussian with FWHM of $\sim 40~\mu{\rm as}$; for reference, the red circle has a diameter of $\sim 50~\mu{\rm as}$, comparable to the expected shadow size for \sgra. Because the angular broadening is significantly smaller than the angular size of the source, the images are not significantly anisotropic, although refractive substructure is apparent in all cases and is strongest for the boxcar model. 
}
  \label{fig:images_rin_Pphi}
\end{center}
\end{figure*}

We also explored the effects of varying the scattering model by simulating images of the scattering on Gaussian sources. To perform these simulations, we used the \texttt{stochastic optics} module of the \texttt{eht-imaging}\footnote{\url{https://github.com/achael/eht-imaging} \citep{Chael_2016}.} software library, as is described in \citet{Johnson2016}. 

Figure~\ref{fig:images_rin_Pphi} shows simulated images of a Gaussian source with FWHM $=8GM/c^2 \approx 40\,\mu{\rm as}$ \citep[i.e., the size of \sgra\ from EHT observations][]{Doeleman2008}. The scattering parameters were chosen to give scattering that matches observations of \sgra\ at centimeter wavelengths \citep{Bower2006}. The panels show three choices of $\rin$ and three models of the angular power distribution. At 1.3\,mm, the introduced anisotropy in the scattered images is hardly noticeable because the source size is significantly larger than the scattering kernel. However, the refractive substructure in the image is quite pronounced and becomes stronger with increasing $\rin$. Refractive fluctuations appear qualitatively different for the boxcar model, relative to the other two models. 

Figure~\ref{fig:images_rin_Pphi} illustrates the two primary effects of scattering on the image: an overall broadening of the image (``blurring'') by the ensemble-average scattering kernel, and the addition of substructure and distortions on the broadened image. Although image blurring by the ensemble-average scattering kernel could be eliminated using deconvolution schemes such as those described in \citet{Fish2014}, refractive substructure will still remain. Eliminating this will require a more sophisticated scheme such as the stochastic optics method described in \citet{Johnson2016}.

\begin{figure*}
\begin{center}
\includegraphics[height=17cm,width=18cm]{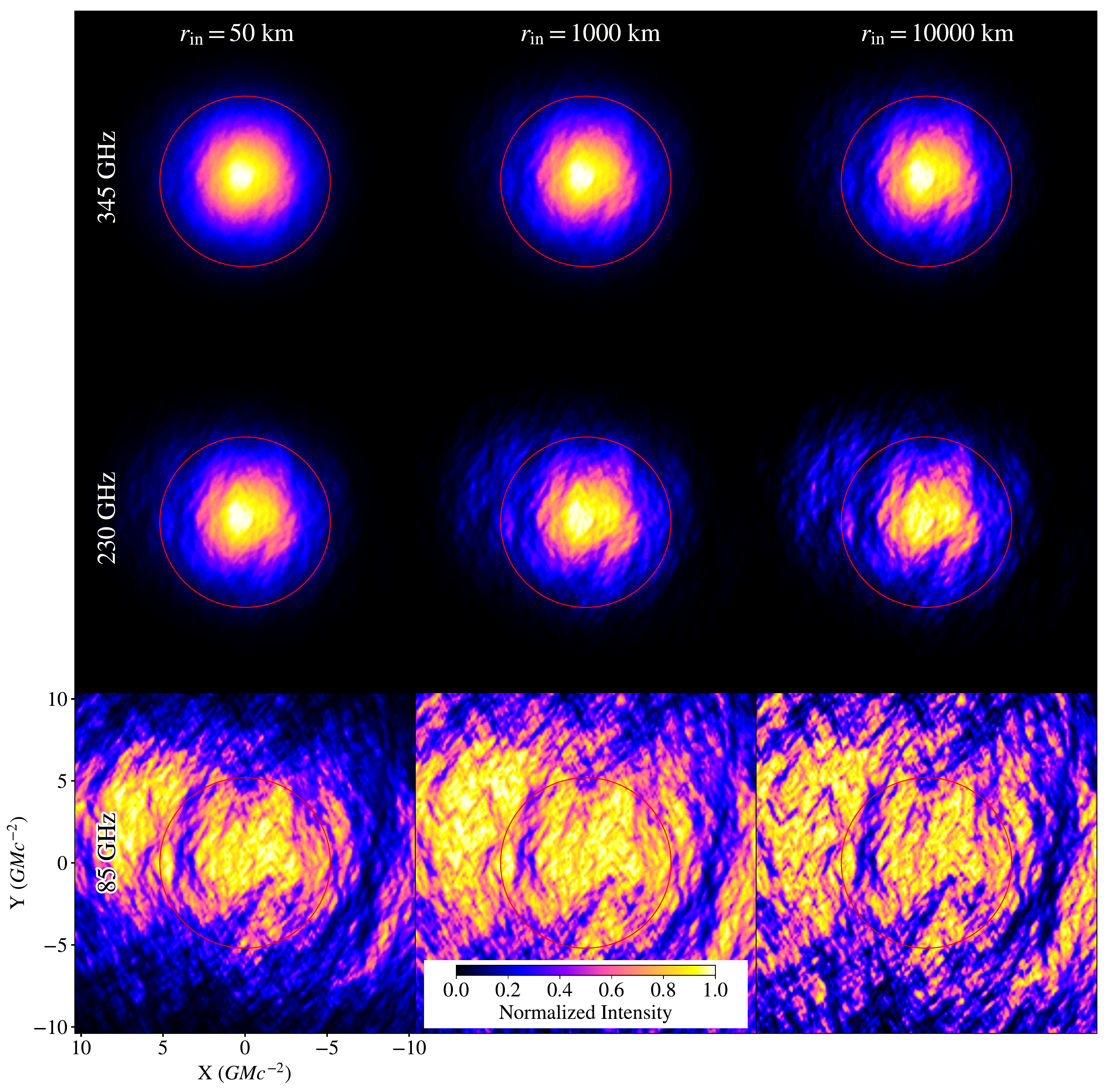}
\caption{Similar to Figure~\ref{fig:images_rin_Pphi} but with rows showing scattering at 85\,GHz ($\sim$3\,mm), 230\,GHz ($\sim$1.3\,mm), and 345\,GHz ($\sim$0.8\,mm) (here we use the von Mises scattering model in all cases). We again use a $\sim 40\,\mu{\rm as}$ Gaussian as the unscattered source for each case, and the effects of scattering-induced anisotropy are evident at $\lambda=$3\,mm. Notably, both image blurring and refractive substructure are insignificant at $\lambda=$3\,mm, regardless of the inner scale.
}
  \label{fig:images_rin_lambda}
\end{center}
\end{figure*}

\begin{figure*}[p]
\begin{center}
\includegraphics[width=\textwidth]{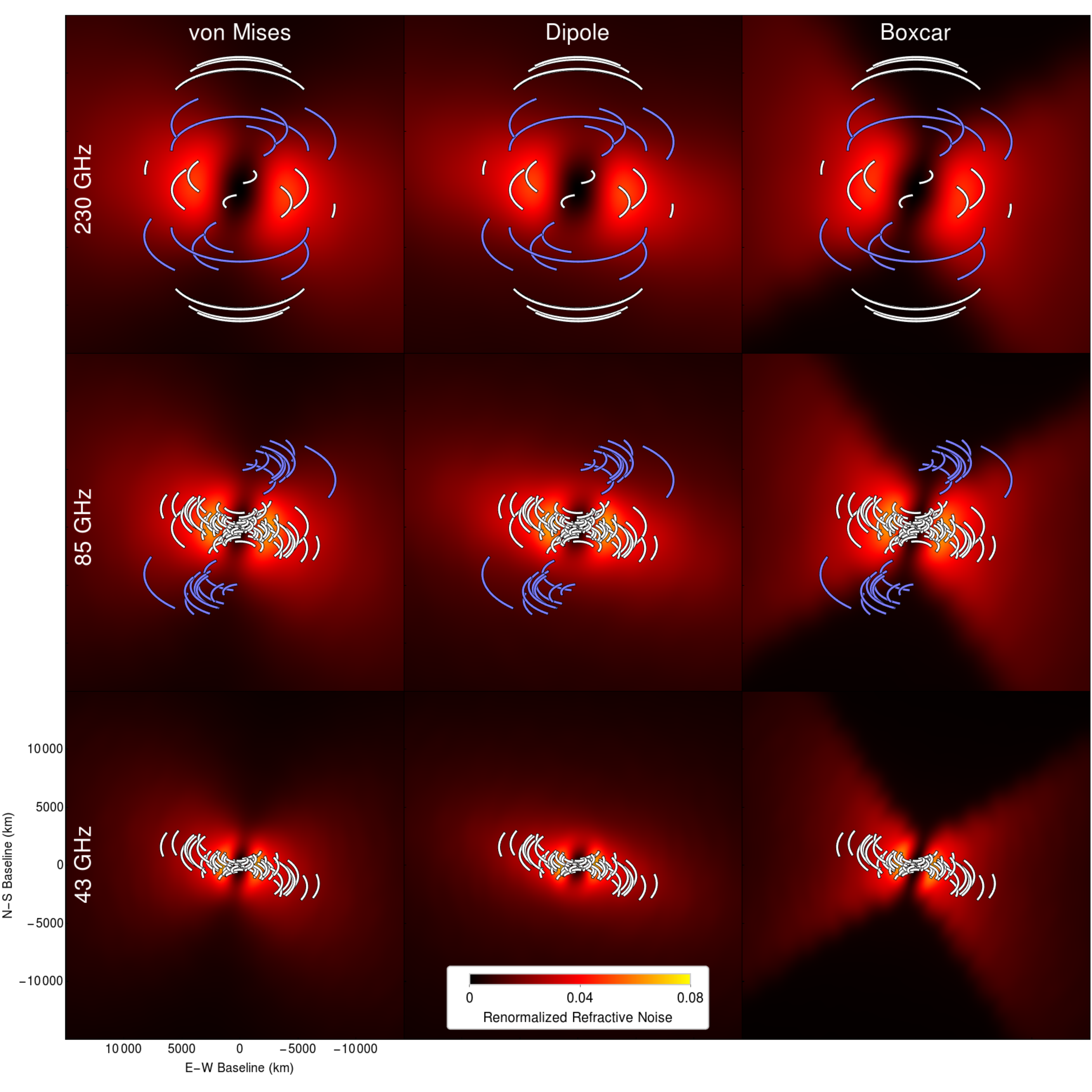}
\caption{The standard deviation of the renormalized refractive noise (i.e., removing contributions from flux modulation and image wander and expressed as a fraction of the total flux density) as a function of the interferometric baseline. Scattering and source parameters are appropriate for \sgra\ (see text for details). Three scattering models are shown for three standard observing frequencies: 230\,GHz (the EHT), 85\,GHz (the GMVA), and 43\,GHz (the VLBA). White tracks show baseline coverage for each array; blue tracks show baselines to ALMA.
}
  \label{fig:refractive_noise_comparison}
\end{center}
\end{figure*}

\begin{figure*}[p]
\begin{center}
  \includegraphics[width=\textwidth]{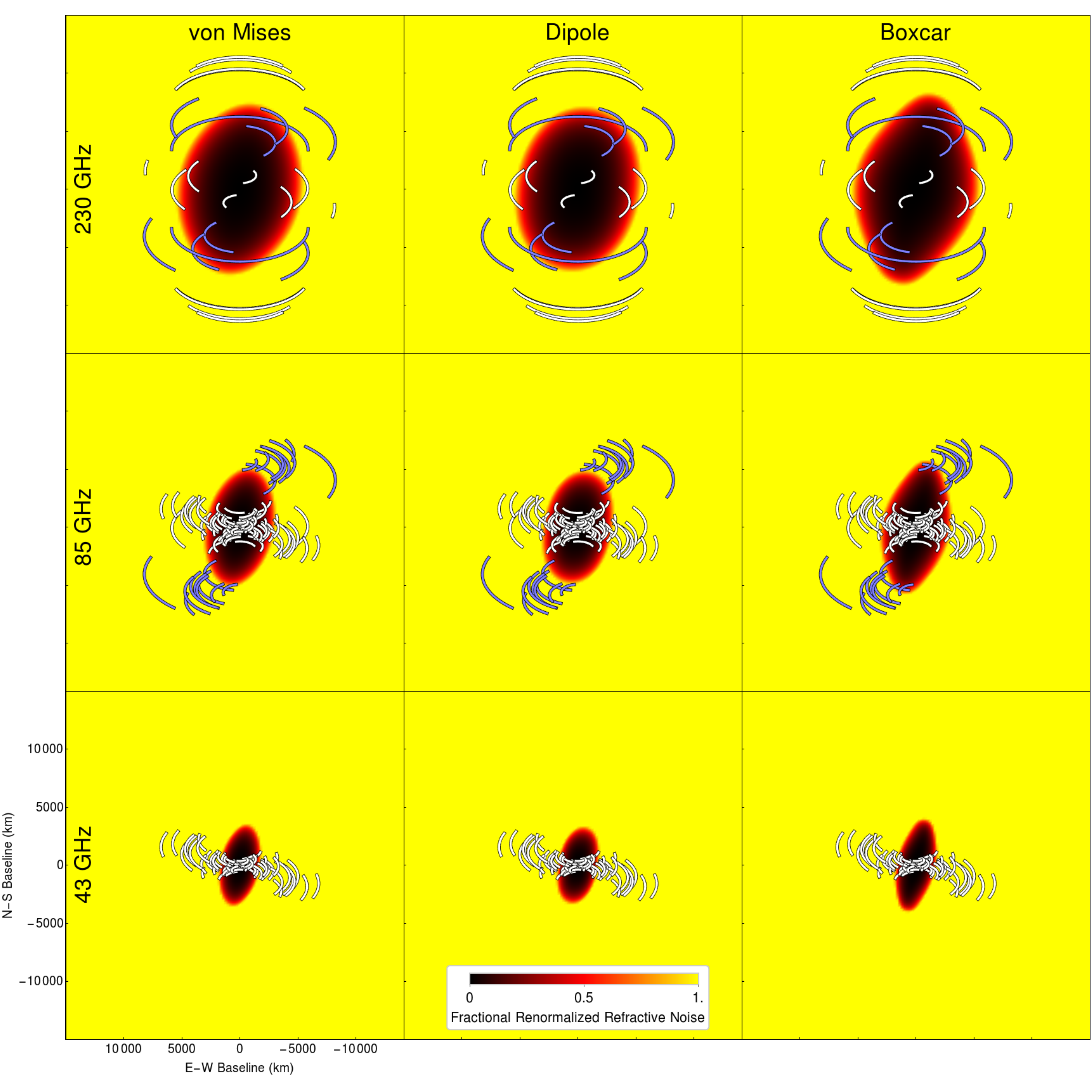}
  \caption{Same as Figure~\ref{fig:refractive_noise_comparison} but showing the renormalized refractive noise on each baseline as a fraction of the ensemble-average visibility on that baseline. Hence, on baselines for which this value is comparable to or greater than unity, measured interferometric visibilities will be dominated by random fluctuations from refractive scattering rather than from intrinsic structure for this Gaussian source model.
}
  \label{fig:fractional_refractive_noise_comparison}
\end{center}
\end{figure*}

Figure \ref{fig:images_rin_lambda} shows the
wavelength dependence of scattering effects, again for three choices of $\rin$, keeping the intrinsic source size the same. Scattering dominates at $\lambda=3$\,mm, to the extent that the anisotropy in the scattering is now quite evident. Refractive effects are also very strong. At 
$\lambda=1.3$\,mm, the case considered in Figure \ref{fig:images_rin_Pphi}, scattering is less dominant, but refractive substructure is still visible. Finally, at $\lambda=0.8$\,mm, scattering is no longer an issue, except perhaps for the largest inner scale we have considered, $\rin=10^4$\,km.

We also used scattering simulations to evaluate the renormalized refractive noise for several cases of interest. Specifically, we considered observations of \sgra\ at 230\, GHz, 85\, GHz, and 43\, GHz. These frequencies represent standard observing frequencies for the EHT, the Global Millimeter VLBI Array (GMVA), and the Very Long Baseline Array (VLBA). At each frequency, we scattered a circular Gaussian source with size given by the approximate inferred size of \sgra\ at that frequency: $40~\mu{\rm as}$ at 230\,GHz \citep[e.g.,][]{Doeleman2008}, $130~\mu{\rm as}$ at 85\,GHz \citep[e.g.,][]{Ortiz2016}, and $300~\mu{\rm as}$ at 43\,GHz \citep[e.g.,][]{Bower2014}. For each frequency and scattering model, we generated 1000 different scattering screens and produced a scattered image for each. We then normalized and centered each image before computing interferometric visibilities. The spread among the complex visibilities on each baseline then gives the renormalized refractive noise. 

Figure~\ref{fig:refractive_noise_comparison} shows the renormalized refractive noise for three scattering models for each of these cases with baseline tracks of the corresponding VLBI arrays overplotted. In this figure, the refractive noise is expressed as the root-mean-square fluctuations divided by the zero-baseline flux density. 
As in Figure~\ref{fig:images_rin_Pphi}, the dipole and amphidirectional von Mises models give comparable results, while the boxcar model gives slightly higher refractive noise. For most baselines, the refractive noise is a few percent of the total image flux density, or roughly ${\sim}100~{\rm mJy}$. If a baseline measures a flux density significantly higher than this value, then the signal must reflect intrinsic source structure.

Figure~\ref{fig:fractional_refractive_noise_comparison} shows the renormalized refractive noise expressed as a fraction of the ensemble-average flux density on the same baseline. These panels thus show the relative dominance of refractive to signal power on each baseline. In these cases, where the unscattered source is a Gaussian, the refractive noise is dominant for most long baselines at each frequency (e.g., all baselines to ALMA are dominated by refractive noise at 3\,mm). However, if the intrinsic source has substructure, then the ensemble-average visibility on long baselines will be higher, decreasing the relative influence of refractive noise on those baselines. 

\section{Conclusions}

In this paper, we developed a model of anisotropic interstellar scattering towards \sgra\ that allows us to extrapolate the properties of the scattering screen, as measured at longer wavelengths, down to the mm wavelengths at which the Event Horizon Telescope will operate when observing \sgra. At these short wavelengths, the diffractive scale in the scattering screen is likely to become comparable to or smaller than the inner scale of turbulence, and this will result in a number of effects: 

\noindent {\em (i)\/} The scattering kernel will evolve from an
anisotropic Gaussian to an anisotropic exponential with a different logarithmic dependence on baseline length.

\noindent {\em (ii)\/} The orientation of anisotropy in the scattering kernel, i.e., the position angle in the sky of the major axis of the scattering kernel, will not change.

\noindent {\em (iii)\/} The degree of anisotropy, i.e., the ratio of the major to the minor axis, will increase with decreasing wavelength in a way that depends on the power-law index $\alpha$ of the turbulence spectrum and on the inner scale $\rin$ of the spectrum of turbulence.

\noindent {\em (iv)\/} The properties of the scattering kernel --- the scatter broadened ensemble average image of a point source --- will depend only weakly on the detailed angular dependence of the power spectrum of density fluctuations in the scattering screen.

\noindent {\em (v)\/} In contrast, the rms refractive visibilities at baseline lengths larger than the diffractive scale will depend very strongly on the angular dependence of the power spectrum of density fluctuations. In some models, the power could be negligibly small at orientations perpendicular to the major axis of the scattering kernel.

\noindent {\em (vi)} Given the expected intrinsic size of \sgra, refractive distortions of the scatter broadened image will be potentially detectable at wavelengths $\lambda \gtrsim 1$\,mm, but will be weak at shorter wavelengths.

Some (but not all) of the above results have been recognized in previous work, though this is the first time these predictions have been collected together in the context of the EHT. In a forthcoming article, we will use archival, high-resolution, long wavelength observations of \sgra\ to infer the model parameters for the scattering screen towards this source, and use the model to make more detailed predictions of the effects of interstellar scattering on imaging observations with the EHT.

\acknowledgements{This was was partially supported by NASA TCAN awards NNX14AB48G and NNX14AB47G as well as by NSF grants AST-1312034, AST-1312651 AST-1440254, and AST-1614868. We also thank the Gordon and Betty Moore Foundation (GBMF-5278) for financial support of this work. DP acknowledges support from the Radcliffe Institute for Advanced Study at Harvard University. L.M. was supported by NFS GRFP grant DGE 1144085. This work was also supported in part by the Black Hole Initiative at Harvard University, which is supported by a grant from the John Templeton Foundation. DP and LM thank the Black Hole Initiative for their hospitality.}

\appendix

\section{Parameters of Approximate Expressions for Model Coefficients}

Tables~(\ref{tab:zeta0})-(\ref{tab:Bmaj}) report the coefficients of the various Pad\'e approximations for the three parameters $\zeta_0$, ${\cal B}_{\rm maj}$, and ${\cal B}_{\rm min}$ of the von Mises, the dipole, and the boxcar models. 

\begin{deluxetable*}{lccc}[t]
\tablewidth{0pt}
\tablecolumns{4}
\tablecaption{Parameters for the Pad\'e approximations for the degree of anisotropy $\zeta_0$\label{tab:zeta0}}
\tablehead{\colhead{Parameters} & \colhead{von Mises Model}
& \colhead{Dipole Model} & \colhead{Boxcar Model}}
\startdata
$\zeta_{u,0}$ & 0           & 0 &   0\\
$\zeta_{u,1}$ & 0           & $2+\alpha$ &   1\\
$\zeta_{u,2}$ & 0.125       & $3(1+\alpha/2)$ & 4\\
$\zeta_{u,3}$ & 0           & $(320+164\alpha+4\alpha^2+\alpha^3)/256$ & 4.69\\
$\zeta_{u,4}$ & 0.00595     & $(64+36\alpha+4\alpha^2+\alpha^3)/512$  & 1.38\\
$\zeta_{u,5}$ & 0           & 0 & 0\\
$\zeta_{u,6}$ & 0.0000465   & 0 & 0\\
$\zeta_{d,0}$ & 1           & 8 & 1 \\
$\zeta_{d,1}$ & 0           & 16 & 4\\
$\zeta_{d,2}$ & 0.214       & $3(112+2\alpha+\alpha^2)/32$ & 6.34\\
$\zeta_{d,3}$ & 0           & $(80+6\alpha+3\alpha^2)/32$ & 4.67\\
$\zeta_{d,4}$ & 0.00744     & $(1920+464\alpha+236\alpha^2+4\alpha^3+\alpha^4)/12288$ & 1.39\\
$\zeta_{d,5}$ & 0           & 0 & 0\\
$\zeta_{d,6}$ & 0.0000496   & 0 & 0\\
\enddata
\end{deluxetable*}

\begin{deluxetable*}{lccc}[t]
\tablewidth{0pt}
\tablecolumns{4}
\tablecaption{Parameters for the Pad\'e approximations for the quantity ${\cal B}_{\rm maj}$\label{tab:Bmaj}}
\tablehead{\colhead{Parameters} & \colhead{von Mises Model}
& \colhead{Dipole Model} & \colhead{Boxcar Model}}
\startdata
${\cal B}_{u,0}$ & 1           & 1 &   1\\
${\cal B}_{u,1}$ & 0           & 2.26 &   5.25\\
${\cal B}_{u,2}$ & 0.321           & 1.74 & 11.64\\
${\cal B}_{u,3}$ & 0           & 0.520 & 11.20\\
${\cal B}_{u,4}$ & 0.01211      & 0.0467  & 4.199\\
${\cal B}_{u,5}$ & 0           & 0 & 0\\
${\cal B}_{u,6}$ & 0.0000795    & 0 & 0\\
${\cal B}_{d,0}$ & 1           & 1 & 1 \\
${\cal B}_{d,1}$ & 0           & 2.32 & 5.23\\
${\cal B}_{d,2}$ & 0.339       & 1.84 & 11.99\\
${\cal B}_{d,3}$ & 0           & 0.564 & 12.47\\
${\cal B}_{d,4}$ & 0.01315     & 0.0519 & 4.672\\
${\cal B}_{d,5}$ & 0           & 0 & 0\\
${\cal B}_{d,6}$ & 0.0000878   & 0 & 0\\
\enddata
\end{deluxetable*}

\begin{deluxetable*}{lccc}[t]
\tablewidth{0pt}
\tablecolumns{4}
\tablecaption{Parameters for the Pad\'e approximations for the quantity ${\cal B}_{\rm min}$\label{tab:Bmin}}
\tablehead{\colhead{Parameters} & \colhead{von Mises Model}
& \colhead{Dipole Model} & \colhead{Boxcar Model}}
\startdata
${\cal M}_{u,0}$ & 1                       & 1 &   1\\
${\cal M}_{u,1}$ & 0                       & 1.738 &   2.486\\
${\cal M}_{u,2}$ & 0.109                   & 0.9567 & 2.254\\
${\cal M}_{u,3}$ & 0                       & 0.1795 & 0.7255\\
${\cal M}_{u,4}$ & 0.00234                 & 0.007659  & 0.0508\\
${\cal M}_{u,5}$ & 0                       & 0 & 0\\
${\cal M}_{u,6}$ & 9.307$\times 10^{-6}$   & 0 & 0\\
${\cal M}_{d,0}$ & 1                       & 1 & 1 \\
${\cal M}_{d,1}$ & 0                       & 1.6755 & 2.288\\
${\cal M}_{d,2}$ & 0.0913                  & 0.8676 & 1.585\\
${\cal M}_{d,3}$ & 0                       & 0.1445 & 0.2821\\
${\cal M}_{d,4}$ & 0.00165                 & 0.004417 & 0.00565\\
${\cal M}_{d,5}$ & 0                       & 0 & 0\\
${\cal M}_{d,6}$ & 5.067$\times 10^{-6}$   & 0 & 0\\
\enddata
\end{deluxetable*}

\bibliographystyle{yahapj}

\bibliography{scattering_v1.0}

\begin{thebibliography}{}
\providecommand\natexlab[1]{#1}
\providecommand\JournalTitle[1]{#1}

\bibitem[{{Armstrong} {et~al.}(1981){Armstrong}, {Cordes}, \&
  {Rickett}}]{Armstrong1981}
{Armstrong}, J.~W., {Cordes}, J.~M., \& {Rickett}, B.~J. 1981,
  \href{http://dx.doi.org/10.1038/291561a0}{\JournalTitle{\nat}, 291, 561}

\bibitem[{{Armstrong} {et~al.}(1995){Armstrong}, {Rickett}, \&
  {Spangler}}]{Armstrong1995}
{Armstrong}, J.~W., {Rickett}, B.~J., \& {Spangler}, S.~R. 1995,
  \href{http://dx.doi.org/10.1086/175515}{\JournalTitle{\apj}, 443, 209}

\bibitem[{{Bardeen}(1973)}]{Bardeen1973}
{Bardeen}, J.~M. 1973, in Black Holes (Les Astres Occlus), ed. C.~{Dewitt} \&
  B.~S. {Dewitt}, 215

\bibitem[{{Blandford} \& {Narayan}(1985)}]{Blandford1985}
{Blandford}, R., \& {Narayan}, R. 1985,
  \href{http://dx.doi.org/10.1093/mnras/213.3.591}{\JournalTitle{\mnras}, 213,
  591}

\bibitem[{{Bower} {et~al.}(2004){Bower}, {Falcke}, {Herrnstein}, {Zhao},
  {Goss}, \& {Backer}}]{Bower2004}
{Bower}, G.~C., {Falcke}, H., {Herrnstein}, R.~M., {et~al.} 2004,
  \href{http://dx.doi.org/10.1126/science.1094023}{\JournalTitle{Science}, 304,
  704}

\bibitem[{{Bower} {et~al.}(2006){Bower}, {Goss}, {Falcke}, {Backer}, \&
  {Lithwick}}]{Bower2006}
{Bower}, G.~C., {Goss}, W.~M., {Falcke}, H., {Backer}, D.~C., \& {Lithwick}, Y.
  2006, \href{http://dx.doi.org/10.1086/508019}{\JournalTitle{\apjl}, 648,
  L127}

\bibitem[{{Bower} {et~al.}(2014{\natexlab{a}}){Bower}, {Deller}, {Demorest},
  {Brunthaler}, {Eatough}, {Falcke}, {Kramer}, {Lee}, \&
  {Spitler}}]{Bower2014a}
{Bower}, G.~C., {Deller}, A., {Demorest}, P., {et~al.} 2014{\natexlab{a}},
  \href{http://dx.doi.org/10.1088/2041-8205/780/1/L2}{\JournalTitle{\apjl},
  780, L2}

\bibitem[{{Bower} {et~al.}(2014{\natexlab{b}}){Bower}, {Markoff}, {Brunthaler},
  {Law}, {Falcke}, {Maitra}, {Clavel}, {Goldwurm}, {Morris}, {Witzel}, {Meyer},
  \& {Ghez}}]{Bower2014}
{Bower}, G.~C., {Markoff}, S., {Brunthaler}, A., {et~al.} 2014{\natexlab{b}},
  \href{http://dx.doi.org/10.1088/0004-637X/790/1/1}{\JournalTitle{\apj}, 790,
  1}

\bibitem[{{Broderick} {et~al.}(2009){Broderick}, {Fish}, {Doeleman}, \&
  {Loeb}}]{Broderick2009}
{Broderick}, A.~E., {Fish}, V.~L., {Doeleman}, S.~S., \& {Loeb}, A. 2009,
  \href{http://dx.doi.org/10.1088/0004-637X/697/1/45}{\JournalTitle{\apj}, 697,
  45}

\bibitem[{{Chael} {et~al.}(2016){Chael}, {Johnson}, {Narayan}, {Doeleman},
  {Wardle}, \& {Bouman}}]{Chael_2016}
{Chael}, A.~A., {Johnson}, M.~D., {Narayan}, R., {et~al.} 2016,
  \href{http://dx.doi.org/10.3847/0004-637X/829/1/11}{\JournalTitle{\apj}, 829,
  11}

\bibitem[{{Chandran} \& {Backer}(2002)}]{Chandran2002}
{Chandran}, B.~D.~G., \& {Backer}, D.~C. 2002,
  \href{http://dx.doi.org/10.1086/340792}{\JournalTitle{\apj}, 576, 176}

\bibitem[{{Cohen} \& {Cronyn}(1974)}]{Cohen1974}
{Cohen}, M.~H., \& {Cronyn}, W.~M. 1974,
  \href{http://dx.doi.org/10.1086/153050}{\JournalTitle{\apj}, 192, 193}

\bibitem[{{Coles} {et~al.}(1987){Coles}, {Rickett}, {Codona}, \&
  {Frehlich}}]{Coles1987}
{Coles}, W.~A., {Rickett}, B.~J., {Codona}, J.~L., \& {Frehlich}, R.~G. 1987,
  \href{http://dx.doi.org/10.1086/165168}{\JournalTitle{\apj}, 315, 666}

\bibitem[{{Cronyn}(1972)}]{Cronyn1972}
{Cronyn}, W.~M. 1972,
  \href{http://dx.doi.org/10.1086/151480}{\JournalTitle{\apj}, 174, 181}

\bibitem[{{Dexter} {et~al.}(2009){Dexter}, {Agol}, \& {Fragile}}]{Dexter2009}
{Dexter}, J., {Agol}, E., \& {Fragile}, P.~C. 2009,
  \href{http://dx.doi.org/10.1088/0004-637X/703/2/L142}{\JournalTitle{\apjl},
  703, L142}

\bibitem[{{Dexter} {et~al.}(2010){Dexter}, {Agol}, {Fragile}, \&
  {McKinney}}]{Dexter2010}
{Dexter}, J., {Agol}, E., {Fragile}, P.~C., \& {McKinney}, J.~C. 2010,
  \href{http://dx.doi.org/10.1088/0004-637X/717/2/1092}{\JournalTitle{\apj},
  717, 1092}

\bibitem[{{Dexter} {et~al.}(2017){Dexter}, {Deller}, {Bower}, {Demorest},
  {Kramer}, {Stappers}, {Lyne}, {Kerr}, {Spitler}, {Psaltis}, {Johnson}, \&
  {Narayan}}]{Dexter2017}
{Dexter}, J., {Deller}, A., {Bower}, G.~C., {et~al.} 2017,
  \href{http://dx.doi.org/10.1093/mnras/stx1777}{\JournalTitle{\mnras}, 471,
  3563}

\bibitem[{{Doeleman} {et~al.}(2001){Doeleman}, {Shen}, {Rogers}, {Bower},
  {Wright}, {Zhao}, {Backer}, {Crowley}, {Freund}, {Ho}, {Lo}, \&
  {Woody}}]{Doeleman2001}
{Doeleman}, S.~S., {Shen}, Z.-Q., {Rogers}, A.~E.~E., {et~al.} 2001,
  \href{http://dx.doi.org/10.1086/320376}{\JournalTitle{\aj}, 121, 2610}

\bibitem[{{Doeleman} {et~al.}(2008){Doeleman}, {Weintroub}, {Rogers},
  {Plambeck}, {Freund}, {Tilanus}, {Friberg}, {Ziurys}, {Moran}, {Corey},
  {Young}, {Smythe}, {Titus}, {Marrone}, {Cappallo}, {Bock}, {Bower},
  {Chamberlin}, {Davis}, {Krichbaum}, {Lamb}, {Maness}, {Niell}, {Roy},
  {Strittmatter}, {Werthimer}, {Whitney}, \& {Woody}}]{Doeleman2008}
{Doeleman}, S.~S., {Weintroub}, J., {Rogers}, A.~E.~E., {et~al.} 2008,
  \href{http://dx.doi.org/10.1038/nature07245}{\JournalTitle{\nat}, 455, 78}

\bibitem[{{Falcke} {et~al.}(2000){Falcke}, {Melia}, \& {Agol}}]{Falcke2000}
{Falcke}, H., {Melia}, F., \& {Agol}, E. 2000,
  \href{http://dx.doi.org/10.1086/312423}{\JournalTitle{\apjl}, 528, L13}

\bibitem[{{Fish} {et~al.}(2011){Fish}, {Doeleman}, {Beaudoin}, {Blundell},
  {Bolin}, {Bower}, {Chamberlin}, {Freund}, {Friberg}, {Gurwell}, {Honma},
  {Inoue}, {Krichbaum}, {Lamb}, {Marrone}, {Moran}, {Oyama}, {Plambeck},
  {Primiani}, {Rogers}, {Smythe}, {SooHoo}, {Strittmatter}, {Tilanus}, {Titus},
  {Weintroub}, {Wright}, {Woody}, {Young}, \& {Ziurys}}]{Fish_2011}
{Fish}, V.~L., {Doeleman}, S.~S., {Beaudoin}, C., {et~al.} 2011,
  \href{http://dx.doi.org/10.1088/2041-8205/727/2/L36}{\JournalTitle{\apjl},
  727, L36}

\bibitem[{{Fish} {et~al.}(2014){Fish}, {Johnson}, {Lu}, {Doeleman}, {Bouman},
  {Zoran}, {Freeman}, {Psaltis}, {Narayan}, {Pankratius}, {Broderick}, {Gwinn},
  \& {Vertatschitsch}}]{Fish2014}
{Fish}, V.~L., {Johnson}, M.~D., {Lu}, R.-S., {et~al.} 2014,
  \href{http://dx.doi.org/10.1088/0004-637X/795/2/134}{\JournalTitle{\apj},
  795, 134}

\bibitem[{{Fish} {et~al.}(2016){Fish}, {Johnson}, {Doeleman}, {Broderick},
  {Psaltis}, {Lu}, {Akiyama}, {Alef}, {Algaba}, {Asada}, {Beaudoin},
  {Bertarini}, {Blackburn}, {Blundell}, {Bower}, {Brinkerink}, {Cappallo},
  {Chael}, {Chamberlin}, {Chan}, {Crew}, {Dexter}, {Dexter}, {Dzib}, {Falcke},
  {Freund}, {Friberg}, {Greer}, {Gurwell}, {Ho}, {Honma}, {Inoue}, {Johannsen},
  {Kim}, {Krichbaum}, {Lamb}, {Le{\'o}n-Tavares}, {Loeb}, {Loinard},
  {MacMahon}, {Marrone}, {Moran}, {Mo{\'s}cibrodzka}, {Ortiz-Le{\'o}n},
  {Oyama}, {{\"O}zel}, {Plambeck}, {Pradel}, {Primiani}, {Rogers}, {Rosenfeld},
  {Rottmann}, {Roy}, {Ruszczyk}, {Smythe}, {SooHoo}, {Spilker}, {Stone},
  {Strittmatter}, {Tilanus}, {Titus}, {Vertatschitsch}, {Wagner}, {Wardle},
  {Weintroub}, {Woody}, {Wright}, {Yamaguchi}, {Young}, {Young}, {Zensus}, \&
  {Ziurys}}]{Fish2016}
{Fish}, V.~L., {Johnson}, M.~D., {Doeleman}, S.~S., {et~al.} 2016,
  \href{http://dx.doi.org/10.3847/0004-637X/820/2/90}{\JournalTitle{\apj}, 820,
  90}

\bibitem[{{Frail} {et~al.}(1994){Frail}, {Diamond}, {Cordes}, \& {van
  Langevelde}}]{Frail1994}
{Frail}, D.~A., {Diamond}, P.~J., {Cordes}, J.~M., \& {van Langevelde}, H.~J.
  1994, \href{http://dx.doi.org/10.1086/187360}{\JournalTitle{\apjl}, 427, L43}

\bibitem[{{Goldreich} \& {Sridhar}(1995)}]{Goldreich1995}
{Goldreich}, P., \& {Sridhar}, S. 1995,
  \href{http://dx.doi.org/10.1086/175121}{\JournalTitle{\apj}, 438, 763}

\bibitem[{{Goldreich} \& {Sridhar}(1997)}]{Goldreich1997}
---. 1997, \href{http://dx.doi.org/10.1086/304442}{\JournalTitle{\apj}, 485,
  680}

\bibitem[{{Goodman} \& {Narayan}(1985)}]{Goodman1985}
{Goodman}, J., \& {Narayan}, R. 1985,
  \href{http://dx.doi.org/10.1093/mnras/214.4.519}{\JournalTitle{\mnras}, 214,
  519}

\bibitem[{{Goodman} \& {Narayan}(1989)}]{Goodman1989}
---. 1989,
  \href{http://dx.doi.org/10.1093/mnras/238.3.995}{\JournalTitle{\mnras}, 238,
  995}

\bibitem[{{Gwinn} {et~al.}(2014){Gwinn}, {Kovalev}, {Johnson}, \&
  {Soglasnov}}]{Gwinn2014}
{Gwinn}, C.~R., {Kovalev}, Y.~Y., {Johnson}, M.~D., \& {Soglasnov}, V.~A. 2014,
  \href{http://dx.doi.org/10.1088/2041-8205/794/1/L14}{\JournalTitle{\apjl},
  794, L14}

\bibitem[{{Johnson}(2016)}]{Johnson2016}
{Johnson}, M.~D. 2016,
  \href{http://dx.doi.org/10.3847/1538-4357/833/1/74}{\JournalTitle{\apj}, 833,
  74}

\bibitem[{{Johnson} \& {Gwinn}(2015)}]{Johnson2015}
{Johnson}, M.~D., \& {Gwinn}, C.~R. 2015,
  \href{http://dx.doi.org/10.1088/0004-637X/805/2/180}{\JournalTitle{\apj},
  805, 180}

\bibitem[{{Johnson} \& {Narayan}(2016)}]{Johnson2016a}
{Johnson}, M.~D., \& {Narayan}, R. 2016,
  \href{http://dx.doi.org/10.3847/0004-637X/826/2/170}{\JournalTitle{\apj},
  826, 170}

\bibitem[{{Johnson} {et~al.}(2015){Johnson}, {Fish}, {Doeleman}, {Marrone},
  {Plambeck}, {Wardle}, {Akiyama}, {Asada}, {Beaudoin}, {Blackburn},
  {Blundell}, {Bower}, {Brinkerink}, {Broderick}, {Cappallo}, {Chael}, {Crew},
  {Dexter}, {Dexter}, {Freund}, {Friberg}, {Gold}, {Gurwell}, {Ho}, {Honma},
  {Inoue}, {Kosowsky}, {Krichbaum}, {Lamb}, {Loeb}, {Lu}, {MacMahon},
  {McKinney}, {Moran}, {Narayan}, {Primiani}, {Psaltis}, {Rogers}, {Rosenfeld},
  {SooHoo}, {Tilanus}, {Titus}, {Vertatschitsch}, {Weintroub}, {Wright},
  {Young}, {Zensus}, \& {Ziurys}}]{Johnson_2015b}
{Johnson}, M.~D., {Fish}, V.~L., {Doeleman}, S.~S., {et~al.} 2015,
  \href{http://dx.doi.org/10.1126/science.aac7087}{\JournalTitle{Science}, 350,
  1242}

\bibitem[{{Kim} {et~al.}(2016){Kim}, {Marrone}, {Chan}, {Medeiros}, {{\"O}zel},
  \& {Psaltis}}]{Kim2016}
{Kim}, J., {Marrone}, D.~P., {Chan}, C.-K., {et~al.} 2016,
  \href{http://dx.doi.org/10.3847/0004-637X/832/2/156}{\JournalTitle{\apj},
  832, 156}

\bibitem[{{Krichbaum} {et~al.}(1998){Krichbaum}, {Graham}, {Witzel}, {Greve},
  {Wink}, {Grewing}, {Colomer}, {de Vicente}, {Gomez-Gonzalez}, {Baudry}, \&
  {Zensus}}]{Krichbaum1998}
{Krichbaum}, T.~P., {Graham}, D.~A., {Witzel}, A., {et~al.} 1998,
  \JournalTitle{\aap}, 335, L106

\bibitem[{{Lazio} \& {Cordes}(1998{\natexlab{a}})}]{Lazio1998a}
{Lazio}, T.~J.~W., \& {Cordes}, J.~M. 1998{\natexlab{a}},
  \href{http://dx.doi.org/10.1086/313129}{\JournalTitle{\apjs}, 118, 201}

\bibitem[{{Lazio} \& {Cordes}(1998{\natexlab{b}})}]{Lazio1998b}
---. 1998{\natexlab{b}},
  \href{http://dx.doi.org/10.1086/306174}{\JournalTitle{\apj}, 505, 715}

\bibitem[{{Lee} \& {Jokipii}(1975{\natexlab{a}})}]{Lee1975b}
{Lee}, L.~C., \& {Jokipii}, J.~R. 1975{\natexlab{a}},
  \href{http://dx.doi.org/10.1086/153458}{\JournalTitle{\apj}, 196, 695}

\bibitem[{{Lee} \& {Jokipii}(1975{\natexlab{b}})}]{Lee1975a}
---. 1975{\natexlab{b}},
  \href{http://dx.doi.org/10.1086/153916}{\JournalTitle{\apj}, 201, 532}

\bibitem[{{Lee} \& {Jokipii}(1975{\natexlab{c}})}]{Lee1975}
---. 1975{\natexlab{c}},
  \href{http://dx.doi.org/10.1086/153994}{\JournalTitle{\apj}, 202, 439}

\bibitem[{{Lo} {et~al.}(1993){Lo}, {Backer}, {Kellermann}, {Reid}, {Zhao},
  {Goss}, \& {Moran}}]{Lo1993}
{Lo}, K.~Y., {Backer}, D.~C., {Kellermann}, K.~I., {et~al.} 1993,
  \href{http://dx.doi.org/10.1038/362038a0}{\JournalTitle{\nat}, 362, 38}

\bibitem[{{Lo} {et~al.}(1998){Lo}, {Shen}, {Zhao}, \& {Ho}}]{Lo1998}
{Lo}, K.~Y., {Shen}, Z.-Q., {Zhao}, J.-H., \& {Ho}, P.~T.~P. 1998,
  \href{http://dx.doi.org/10.1086/311726}{\JournalTitle{\apjl}, 508, L61}

\bibitem[{{Lu} {et~al.}(2011){Lu}, {Krichbaum}, {Eckart}, {K{\"o}nig},
  {Kunneriath}, {Witzel}, {Witzel}, \& {Zensus}}]{Lu2011}
{Lu}, R.-S., {Krichbaum}, T.~P., {Eckart}, A., {et~al.} 2011,
  \href{http://dx.doi.org/10.1051/0004-6361/200913807}{\JournalTitle{\aap},
  525, A76}

\bibitem[{{Luminet}(1979)}]{Luminet1979}
{Luminet}, J.-P. 1979, \JournalTitle{\aap}, 75, 228

\bibitem[{{Mardia} \& {Jupp}(1999)}]{Mardia1999}
{Mardia}, K., \& {Jupp}, P. 1999

\bibitem[{{Mo{\'s}cibrodzka} {et~al.}(2014){Mo{\'s}cibrodzka}, {Falcke},
  {Shiokawa}, \& {Gammie}}]{Moscibrodzka2014}
{Mo{\'s}cibrodzka}, M., {Falcke}, H., {Shiokawa}, H., \& {Gammie}, C.~F. 2014,
  \href{http://dx.doi.org/10.1051/0004-6361/201424358}{\JournalTitle{\aap},
  570, A7}

\bibitem[{{Mo{\'s}cibrodzka} {et~al.}(2009){Mo{\'s}cibrodzka}, {Gammie},
  {Dolence}, {Shiokawa}, \& {Leung}}]{Moscibrodzka2009}
{Mo{\'s}cibrodzka}, M., {Gammie}, C.~F., {Dolence}, J.~C., {Shiokawa}, H., \&
  {Leung}, P.~K. 2009,
  \href{http://dx.doi.org/10.1088/0004-637X/706/1/497}{\JournalTitle{\apj},
  706, 497}

\bibitem[{{Narayan}(1992)}]{Narayan1992}
{Narayan}, R. 1992,
  \href{http://dx.doi.org/10.1098/rsta.1992.0090}{\JournalTitle{Philosophical
  Transactions of the Royal Society of London Series A}, 341, 151}

\bibitem[{{Narayan} \& {Goodman}(1989)}]{Narayan1989a}
{Narayan}, R., \& {Goodman}, J. 1989,
  \href{http://dx.doi.org/10.1093/mnras/238.3.963}{\JournalTitle{\mnras}, 238,
  963}

\bibitem[{{Narayan} \& {Hubbard}(1988)}]{Narayan1988}
{Narayan}, R., \& {Hubbard}, W.~B. 1988,
  \href{http://dx.doi.org/10.1086/166020}{\JournalTitle{\apj}, 325, 503}

\bibitem[{{Ortiz-Le{\'o}n} {et~al.}(2016){Ortiz-Le{\'o}n}, {Johnson},
  {Doeleman}, {Blackburn}, {Fish}, {Loinard}, {Reid}, {Castillo}, {Chael},
  {Hern{\'a}ndez-G{\'o}mez}, {Hughes}, {Le{\'o}n-Tavares}, {Lu}, {Monta{\~n}a},
  {Narayanan}, {Rosenfeld}, {S{\'a}nchez}, {Schloerb}, {Shen}, {Shiokawa},
  {SooHoo}, \& {Vertatschitsch}}]{Ortiz2016}
{Ortiz-Le{\'o}n}, G.~N., {Johnson}, M.~D., {Doeleman}, S.~S., {et~al.} 2016,
  \href{http://dx.doi.org/10.3847/0004-637X/824/1/40}{\JournalTitle{\apj}, 824,
  40}

\bibitem[{{{\"O}zel} {et~al.}(2000){{\"O}zel}, {Psaltis}, \&
  {Narayan}}]{Ozel2000}
{{\"O}zel}, F., {Psaltis}, D., \& {Narayan}, R. 2000,
  \href{http://dx.doi.org/10.1086/309396}{\JournalTitle{\apj}, 541, 234}

\bibitem[{{Psaltis} {et~al.}(2015){Psaltis}, {{\"O}zel}, {Chan}, \&
  {Marrone}}]{Psaltis2015}
{Psaltis}, D., {{\"O}zel}, F., {Chan}, C.-K., \& {Marrone}, D.~P. 2015,
  \href{http://dx.doi.org/10.1088/0004-637X/814/2/115}{\JournalTitle{\apj},
  814, 115}

\bibitem[{{Rickett} {et~al.}(2009){Rickett}, {Johnston}, {Tomlinson}, \&
  {Reynolds}}]{Rickett_2009}
{Rickett}, B., {Johnston}, S., {Tomlinson}, T., \& {Reynolds}, J. 2009,
  \href{http://dx.doi.org/10.1111/j.1365-2966.2009.14471.x}{\JournalTitle{\mnras},
  395, 1391}

\bibitem[{{Rickett}(1990)}]{Rickett1990}
{Rickett}, B.~J. 1990,
  \href{http://dx.doi.org/10.1146/annurev.aa.28.090190.003021}{\JournalTitle{\araa},
  28, 561}

\bibitem[{{Shen} {et~al.}(2005){Shen}, {Lo}, {Liang}, {Ho}, \&
  {Zhao}}]{Shen2005}
{Shen}, Z.-Q., {Lo}, K.~Y., {Liang}, M.-C., {Ho}, P.~T.~P., \& {Zhao}, J.-H.
  2005, \href{http://dx.doi.org/10.1038/nature04205}{\JournalTitle{\nat}, 438,
  62}

\bibitem[{{Spangler} \& {Gwinn}(1990)}]{Spangler1990}
{Spangler}, S.~R., \& {Gwinn}, C.~R. 1990,
  \href{http://dx.doi.org/10.1086/185700}{\JournalTitle{\apjl}, 353, L29}

\bibitem[{{Spitler} {et~al.}(2014){Spitler}, {Lee}, {Eatough}, {Kramer},
  {Karuppusamy}, {Bassa}, {Cognard}, {Desvignes}, {Lyne}, {Stappers}, {Bower},
  {Cordes}, {Champion}, \& {Falcke}}]{Spitler2014}
{Spitler}, L.~G., {Lee}, K.~J., {Eatough}, R.~P., {et~al.} 2014,
  \href{http://dx.doi.org/10.1088/2041-8205/780/1/L3}{\JournalTitle{\apjl},
  780, L3}

\bibitem[{{van Langevelde} {et~al.}(1992){van Langevelde}, {Frail}, {Cordes},
  \& {Diamond}}]{vanLangevelde1992}
{van Langevelde}, H.~J., {Frail}, D.~A., {Cordes}, J.~M., \& {Diamond}, P.~J.
  1992, \href{http://dx.doi.org/10.1086/171750}{\JournalTitle{\apj}, 396, 686}

\end{thebibliography}

\end{document}